\documentclass[fleqn,usenatbib]{mnras}

\usepackage[T1]{fontenc}

\DeclareRobustCommand{\VAN}[3]{#2}
\let\VANthebibliography\thebibliography
\def\thebibliography{\DeclareRobustCommand{\VAN}[3]{##3}\VANthebibliography}

\usepackage{graphicx}	
\usepackage{amsmath}	
\usepackage{amssymb}	

\usepackage{txfonts}

\usepackage[dvipsnames]{xcolor}

\usepackage{siunitx}
\usepackage{listings}
\usepackage{todonotes}
\usepackage[normalem]{ulem}

\DeclareSIUnit \h {\ensuremath{\mathit{h}}}
\DeclareSIUnit \parsec {pc}
\DeclareSIUnit \pc {pc}
\DeclareSIUnit \msol {\ensuremath{M_{\odot}}}
\DeclareSIUnit \kpch {\kilo\pc\per\h}
\DeclareSIUnit \Mpch {\mega\pc\per\h}

\newcommand{\rvtext}[1]{#1}
\newcommand{\rvcom}[1]{}
\newcommand{\rvsout}[1]{}

\newcommand{\myvec}[1]{\boldsymbol{#1}}

\newcommand{\TTK}{\textsc{TopologyToolKit} }

\newcommand{\Tid}[0]{\mathbfss{T}}
\newcommand{\phiboost}[0]{\phi_{\rm{boost}}}
\newcommand{\phisad}[0]{\ensuremath{\phi_\mathrm{saddle}}}
\newcommand{\Mtid}[0]{M_{\rm{tid}}}

\newcommand{\subfind}{\textsc{subfind}}

\newcommand{\norm}[1]{\left\lVert#1\right\rVert}

\newcommand{\citncdm}[0]{Stücker et al (in prep. A)}
\newcommand{\citsubhalo}[0]{Stücker et al (in prep. B)}
\newcommand{\citlagrangian}[0]{Stücker et al (in prep. C)}

\title[The Boosted Potential]{The Boosted Potential}

\author[J. St\"ucker et al.]{
Jens St\"ucker,$^{1}$\thanks{E-mail: jstuecker@dipc.org}
Raul E. Angulo$^{1,2}$ and
Philipp Busch$^{3,4}$
\\
$^{1}$Donostia International Physics Centre (DIPC), Paseo Manuel de Lardizabal 4, 20018 Donostia-San Sebastian, Spain.\\
$^{2}$IKERBASQUE, Basque Foundation for Science, E-48013, Bilbao, Spain.\\
$^{3}$Department of Natural Science, The Open University of Israel, 1 University Road, P. O. Box 808, Raanana 43107, Israel\\
$^{4}$Max-Planck-Institut f\"ur Astrophysik, Postfach 1317, D-85741 Garching, Germany
}

\date{Accepted XXX. Received YYY; in original form ZZZ}

\pubyear{2021}

\begin{document}
\label{firstpage}
\pagerange{\pageref{firstpage}--\pageref{lastpage}}
\maketitle

\begin{abstract}
The global gravitational potential, $\phi$, is not commonly employed in the analysis of cosmological simulations, since its level sets do not show any clear correspondence to the underlying density field and its persistent structures. Here, we show that the potential becomes a locally meaningful quantity when considered from a boosted frame of reference, defined by subtracting a uniform gradient term $\phi_{\rm{boost}}(\boldsymbol{x}) = \phi(\boldsymbol{x}) + \boldsymbol{x} \cdot \boldsymbol{a}_0$ with acceleration $\boldsymbol{a}_0$. We study this ``boosted potential'' in a variety of scenarios and propose several applications: (1) The boosted potential can be used to define a binding criterion that naturally incorporates the effect of tidal fields. This solves several problems of commonly-used self-potential binding checks: i) it defines a tidal boundary for each halo, ii) it is much less likely to misidentify caustics as haloes (specially in the context of warm dark matter cosmologies), and iii) performs better at identifying virialized regions of haloes -- yielding to the expected value of 2 for the virial ratio. (2) This binding check can be generalized to filaments and other cosmic structures. (3) The boosted potential facilitates the understanding of the disruption of satellite subhaloes. We propose a picture where most mass loss is explained through a lowering of the escape energy through the tidal field. (4) We discuss the possibility of understanding the topology of the potential field in a way that is independent of constant offsets in the first derivative $\boldsymbol{a}_0$. We foresee that this novel perspective on the potential can help to develop more accurate models and improve our understanding of structure formation. \rvcom{(We shortened some sentences in the abstract to be below 250 words.)}
\end{abstract}

\begin{keywords}
cosmology: theory -- large-scale structure of Universe -- methods: numerical
\end{keywords}


\section{Introduction}
\begin{figure}
    \centering
    \includegraphics[width=\columnwidth]{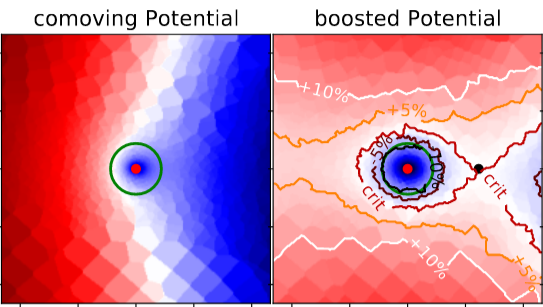}
    \caption{Left the potential of a cosmological simulation when imposing global vacuum boundary conditions (GVBC). Right: the boosted potential which is constructed by subtracting a uniform gradient term. Both panels correspond to the same region. The green circle indicates the virial radius, $R_{\rm{200b}}$, of a halo. The critical energy level is marked red in the right image. Through a ``boost'' operation the potential landscape becomes locally meaningful.}
    \label{fig:global_vs_boosted_pot}
\end{figure}

 In cosmological simulations, the peculiar gravitational potential, $\phi$, is defined through the Poisson equation:
\begin{align}
    \Delta \phi &= 4 \pi G \left( \rho - \rho_0 \right), \label{eqn:poisson}
\end{align}
where $G$ is the gravitational constant, $\rho$ is the mass density in comoving space and $\rho_0$ is the mean density of the universe. \footnote{Note that definitions of the potential may vary by powers of the scale-factor $a$. We follow the notation of \citet{springel_2005} here.} However, since the Poisson equation only restricts the second derivatives of the potential, a choice of boundary conditions is needed to uniquely define $\phi$. Depending on the problem, the chosen boundary conditions are either 'periodic' or 'vacuum' where $\phi \xrightarrow[r \rightarrow \infty]{} 0$ if the density approaches the mean density at infinity $\rho \rightarrow \rho_0$. Here, we will refer to both cases as global vacuum boundary conditions (GVBC), since in both the potential is constant and zero in the absence of density perturbations in the considered volume.

In the computational cosmology community, this potential with GVBC is usually discussed only as an intermediate quantity to define the forces that determine the time evolution of particle trajectories, but it is hardly used in the analysis  of simulations. For example, we do not know of any structure finder based on the global potential. This might seem somewhat surprising, since undergraduate physics books usually teach how potential landscapes help to understand and predict which phase-space regions particles have access to -- and, therefore, which orbits could be considered as bound to different valleys in the potential field.

The reason behind this apparent conflict becomes clear in the left panel of Figure \ref{fig:global_vs_boosted_pot}: Here we display the potential around a large halo in a cosmological simulation. The green circle marks the radius, $R_{\rm{200b}}$, in which the density is $200$ times larger than the mean matter density of the universe. Inside $R_{\rm{200b}}$, the halo is considered as a roughly bound and virialized object. In this figure we see that there is little correspondence between equipotential lines and the boundaries of the structure. This is because the potential is dominated by the largest structures which induce strong gradients even far away. Further, the potential can be strongly time-dependent close to moving objects. In fact, this halo is in a free fall motion towards the positive x-direction and thus the potential minimum moves along with it -- this gives the global potential in the immediate surrounding of the halo a strong time dependence. Therefore, the global potential is typically discarded in favor of the 'self-potential', which considers only the local mass as source of the potential and, thus, discards all  contribution of the surrounding material. 

In this article we propose an alternative locally-meaningful potential landscape: the 'boosted potential'. Large-scale gradients in the potential indicate that mass elements will move with an approximately uniform acceleration  $\myvec{a}_0$. Such large-scale gradients can for example be measured by taking local averages over the acceleration of a group of particles. We can consider the dynamics from a boosted frame of reference, which is defined by the coordinate transformation
\begin{align}
  \myvec{x} \rightarrow \myvec{x} - \myvec{v}_0 t - \frac{1}{2} \myvec{a}_0 t^2  \rm{ .} \label{eqn:boostedpos} 
\end{align}
When switching to this accelerated frame, we have to consider an additional apparent acceleration $\myvec{a}_0$. Therefore, the potential in the boosted frame has the form
\begin{align}
    \phi_{\rm{boost}}(\myvec{x}) &= \phi(\myvec{x}) + \myvec{a}_0 \cdot \myvec{x} \rm{ .} \label{eqn:boostedpot}
\end{align}
We refer to this as ``boosted potential'', $\phi_{\rm{boost}}$ . Note that $\phi_{\rm{boost}}$ is also a solution to the Poisson equation \eqref{eqn:poisson}, but under different boundary conditions. 

We display the boosted potential in the right panel of Figure \ref{fig:global_vs_boosted_pot}. The ``boost'' operation has cancelled out the overall gradient in the region around the halo and, thus, we have created a locally-meaningful potential landscape. The potential in this frame has only a weak explicit time dependence, and as we will see later in this article, the associated energy is approximately conserved. Furthermore, a natural boundary of the halo can be defined as the shallowest level set of the potential that closes around the minimum. This critical level corresponds to an effective escape energy. This can be used to define a criterion for identifying bound objects that properly accounts for the effect of tidal fields. 

In this article, we will discuss this and other aspects in greater detail. We will also present a variety of possible applications of the boosted potential, which we hope will motivate others to incorporate this novel perspective to specific problems. Therefore, the article is structured as follows: 

In Section \ref{sec:math} we will give a more detailed explanation of the boosted potential in the context of the well understood restricted three-body problem. Further, we will explain how this gives rise to the notions of persistence and tidal boundaries. 

In Section \ref{sec:boostedhaloes} we will explore the boosted potential for haloes of a CDM simulation and introduce the boosted potential binding check (BPBC). We show how the BPBC has advantages in comparison to traditional binding checks in selecting bound and virialized regions of a system and in discarding caustics and other unbound overdensities in warm dark matter simulations.

In Section \ref{sec:cosmicweb} we will argue that the language of the boosted potential can also be used to understand the structure of filaments and the cosmic web in general. By using one filament in a cosmological context as an example, we will show that the boosted potential can be used to identify the transition from a filament to a wall-like structure and to define a meaningful binding criterion in slices orthogonal to the filament axis.

In Section \ref{sec:sdisruption} we discuss how the boosted potential can be used to study the disruption of satellite subhaloes within their host halo. This motivates us to propose a ``deforming bowl'' picture of subhalo disruption where mass loss is explained by a lowering of the escape energy caused by the tidal field. This motivates the adiabatic limit of the evolution of a halo in an infinitely slowly applied tidal field as a reasonable model for the asymptotic mass-loss of a subhalo.

Finally, in Section \ref{sec:advancedmath} we will discuss whether it is possible to infer properties like potential valleys and their persistence from the potential field without explicitly moving into a boosted frame, but in a way that is completely independent of an absolute offset in the first derivative $\myvec{a}_0$. We will show that we can define topological properties of the potential field that are independent of $\myvec{a}_0$. However, we have not been able to define an algorithm which effectively infers these in two or more dimensional universes. Therefore, we present an unsolved mathematical problem, that -- if solved -- might help to pave the way for a universal structure finder that only operates with the potential field.

\section{The boosted Potential in the restricted three-body problem} \label{sec:math}

\begin{figure*}
    \centering
    \includegraphics[width=\textwidth]{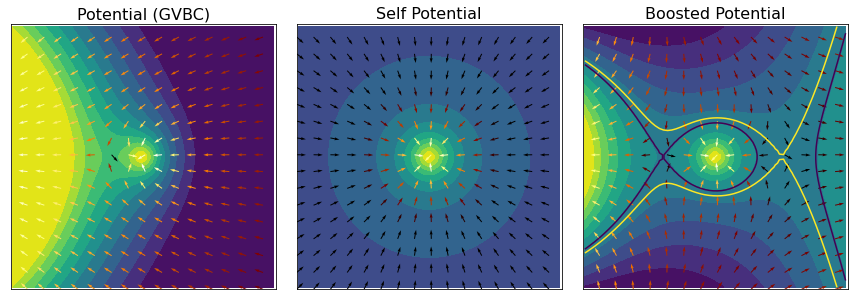}\\
    \includegraphics[width=\textwidth]{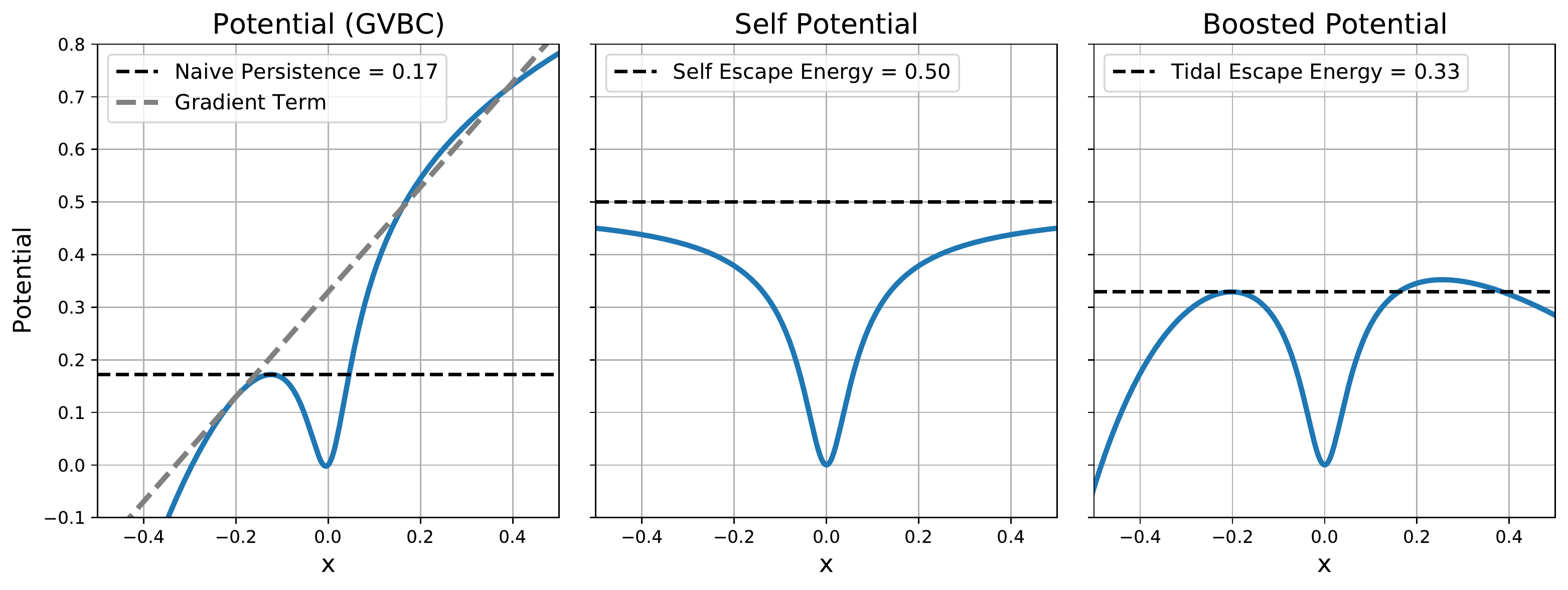}\\
    \includegraphics[width=\textwidth]{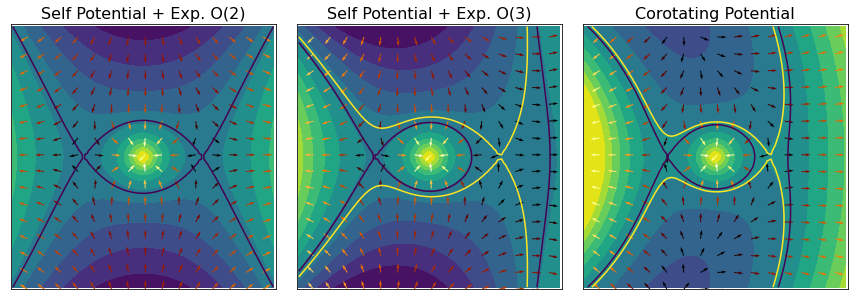}
    \caption{Different potential definitions illustrated in 2-D (top and bottom panels) and in 1-D (central three panels). The potential field comes from two (softened) point masses with a mass ratio of 40:1 with the smaller mass in the center and the larger mass at $x=-1$. Contours show equi-potential lines and vectors show the negative gradient colour-coded by the absolute value of the gradient. Top-left and center-left: the gravitational potential with global vacuum boundary conditions (GVBC). This definition of the potential shows little resemblance with the spatial region which can host bound particles (for the restricted three body problem this is the black contour in the bottom right panel), since it is dominated by a global gradient term. On the other hand, the self-potential (top-center and center-center) would permit particles to be bound which are arbitrarily far away from the center of mass of the object -- so long as if they had a small enough relative velocity. A more meaningful potential landscape is given by the boosted potential (top-right and center-right panels), which has been constructed by subtracting a uniform gradient term from the global potential. The critical energy levels are a good approximation to those of the corotating potential (bottom right). This potential includes the effects of the tidal field -- effectively lowering the ``escape energy'' in the radial (here x-) directions, and increasing it in the opposite directions. The two panels in the bottom left show that the boosted potential can be approximated through the self-potential plus a multipole expansion of the external contributions.}
    \label{fig:twobody}
\end{figure*}

Before we consider the boosted potential in a cosmological context, we will briefly discuss it in the context of the well-studied restricted three-body problem \citep{szebehely_1967, henon_1997, valtonen_2006, BinneyTremaine2008}. This is arguably the simplest problem that motivates the notion of the boosted potential, and we will later see that it can be easily generalized to a cosmological scenario.

\subsection{The restricted three-body problem}

In the restricted tree-body problem, a small point mass $m_b$ is in a circular orbit around a far more massive point mass $m_a$ (e.g., the Sun-Earth system). The potential with global vacuum boundary conditions (GVBC) is given by
\begin{align}
    \phi(\myvec{x}) &= - \frac{G m_a}{|\myvec{x} - \myvec{x}_a|} - \frac{G m_b}{|\myvec{x} - \myvec{x}_b (t)|} \\
             &=: \phi_{\rm{ext}}(\myvec{x}) + \phi_{\rm{self}}(\myvec{x}),
\end{align}
where we assume $\myvec{x}_a$ to be constant for simplicity (which is approximately correct for $m_a \gg m_b$)  and $\myvec{x}_b(t)$ corresponds to a circular motion around $\myvec{x}_a$ with radius $r_b$. The restricted three-body problem then poses the question how a massless test particle moves in this potential landscape. Here, we adopt the perspective of a test particle that is near $\myvec{x}_b$ and, therefore, have labeled the potential contribution from $m_b$ as self-potential $\phi_{\rm{self}}$ and that of $m_a$ as external potential $\phi_{\rm{ext}}$.

We show the global potential in the top- and center-left panels of Figure \ref{fig:twobody}. Of course, the global potential is highly time-dependent since the smaller mass will move over time. We can see that the global potential is dominated by a global gradient term at the location of the substructure. \rvtext{For visual purposes we have used a slightly softened version of the point-mass potential ($\phi \propto (|\myvec{x}| + \epsilon)^{-1}$ with $\epsilon = 0.1$) in this figure.}

\rvtext{One could estimate the depth of the potential well around $\myvec{x}_b$ in this potential landscape by applying a watershed algorithm or using the tools of discrete morse theory \citep{forman_1998, forman_2002}. The result would be the 'persistence' $\Delta \phi = \phi_{\rm{saddle}}  - \phi_{\rm{min}}$ where $\phi_{\rm{min}}$ is the potential value in the minimum and $ \phi_{\rm{saddle}}$ the saddle point of the potential above which equi-potential merge with those that are closing around a deeper minimum. This value of the persistence could be considered as a 'naive' estimate of the escape energy that is needed to escape the potential well from the minimum. However, we will see that this ``naive persistence'' is much lower than the actual energy needed to escape the potential well of $m_b$. In principle, in presence of an even larger gradient the minimum could disappear all-together, but, as we shall see, this would not mean that there cannot be a bound structure.}

\subsection{The self-potential}

In the upper two central panels of Figure \ref{fig:twobody} we show the self-potential $\phi_{\rm{self}}$ of the mass $m_b$. This is the notion of the potential that (to date) is used in all cosmological structure finders which seek to identify a group of particles gravitationally bound -- for example, \subfind{} \citep{springel_2001}, \textsc{rockstar} \citep{behroozi_2013}, \textsc{hbt+} \citep{han_2018} and \textsc{subfind-hbt} \citep{springel_2020}. The self-potential can be constructed by selecting a subset of the mass which shall be considered for a binding check (in this case $m_b$) and solving Poisson's equation for this subset of the mass under the assumption of vacuum boundary conditions. The self-potential is a meaningful description of the potential landscape near $\myvec{x}_b$, and it does not suffer from the effect of the large-scale gradient term like the global potential does. However, in principle it also allows particles arbitrary far from $\myvec{x}_b$ to be considered as bound to $m_b$, and further, it depends on a somewhat arbitrary pre-selection of the mass sourcing the self-potential. In practice, these problems are reduced by imposing a suitable domain-specific pre-selection of particles to be considered for the binding check -- for example, through the friends-of-friends method (FOF) which detects overdensities, or by splitting substructures along saddle points in density \citep{springel_2001} or in phase space \citep{behroozi_2013}.

\subsection{The boosted potential}

Finally, in the right panels of Figure \ref{fig:twobody} we show the boosted potential, $\phi_{\rm{boost}}$, which we have defined by subtracting the gradient-term as in equation \eqref{eqn:boostedpot} with the gradient
\begin{align}
    \myvec{a}_0 &=  - \frac{\partial \phi_{\rm{ext}} }{\partial \myvec{x}}   (\myvec{x}_b) \\
                &= - G m_a  \frac{\myvec{x}_b - \myvec{x}_a}{|\myvec{x}_b  - \myvec{x}_a|^3} .
\end{align}

We can see in Figure \ref{fig:twobody} that the boosted potential has a similar shape to the self-potential near the center of $\myvec{x}_b$ where the effect of the tidal field is negligible. However, further away from the center the shape strongly distorts through the tidal field, leading to a cusp-like shape at the critical contour (marked in black) at the saddle point energy level $\phi_{\rm{saddle}}$. Beyond the critical contour the equipotential lines open up. 

We dub the energy difference $\Delta \phi_{\rm{esc}} = \phi_{\rm{saddle}} - \phi_{\rm{min}}$ (with $\phi_{\rm{min}}$ as the local minimum of the boosted potential around $\myvec{x}_b$) the ``persistence of the minimum'', the ``escape'' energy level or the ``tidal energy''. In the case of point masses, this difference is not well defined since the potential in the minimum goes to negative infinity. However, for typical potentials encountered in cosmology -- like an Navarro–Frenk–White (NFW) profile \citep{navarro_1997} -- the potential is finite in the center. We can see that this ``tidal escape energy'' is lower than that defined through the self-potential. The tidal field presses the potential contours of the self-potential down in one direction, while lifting them up in the other directions. The escape velocity needed to reach infinity from the surface of the earth according to the self-potential is $\sqrt{2 \Delta \phi_\infty} = \SI{11.186}{\kilo \meter \per \second}$. The velocity needed to reach the saddle point in the boosted potential is lower $\sqrt{2 \Delta \phi_{\rm{esc},b}} = \SI{11.155}{\kilo \meter \per \second}$. We shall later see that such differences in escape energies can be important for cosmological structures (Section \ref{sec:boostedhaloes}) and that the lowering of the escape energy can be understood as the main driver of the tidal mass loss of subhaloes (see Section \ref{sec:sdisruption}).

\subsection{Relation between self-potential and boosted potential}

We note that the gradient of the external contributions (of $m_a$) vanishes for the boosted potential at $\myvec{x}_b$ by construction:
\begin{align}
    \phiboost(\myvec{x}) &= \phi_{\rm{self}} (\myvec{x}) + \left( \phi_{\rm{ext}} (\myvec{x}) - (\myvec{x} - \myvec{x}_b) \frac{\partial \phi_{\rm{ext}} }{\partial \myvec{x}} (\myvec{x}_b) \right)
\end{align}
Therefore, a multipole expansion of the potential contributions of $m_a$ around $\myvec{x}_b$ would have the form
\begin{align}
    \phiboost \approx& \phi_{\rm{self}} (\myvec{x}) + \frac{1}{2} \Delta x_i \Delta x_j \partial_{ij} \phi_{\rm{ext}}(\myvec{x}_b)\nonumber \\  
    &+ \frac{1}{6} \Delta x_i \Delta x_j \Delta x_j \partial_{ijk} \phi_{\rm{ext}}(\myvec{x}_b) + ... \label{eqn:tidmultipole}
\end{align}
where $\Delta \myvec{x} = \myvec{x} - \myvec{x}_b$ and we sum over repeated indices and neglected the irrelevant zeroth order offset in the potential. We can identify the second order term with the tidal tensor of the external potential
\begin{align}
    T_{ij} = - \partial_{ij} \phi_{\rm{ext}} (\myvec{x}_b) \rm{.}
\end{align}
Therefore, \emph{the boosted potential corresponds at the lowest order to the self-potential plus the effect of an external tidal field}. Neglecting the effect of external tides is arguably the most important shortcoming of self-potential binding checks. For example, \citet{behroozi_2014} have found that tidal fields significantly affect dark matter haloes that lie well beyond the virial radius of any other halo. The boosted potential appears to us as the most natural way to incorporate the effect of the tidal field and possible higher order contributions.

However, we note that a legitimate approximate way to capture it in the framework of existing structure finders would be through a multipole expansion of the external field contributions (above order two) as in equation \eqref{eqn:tidmultipole}. To demonstrate this, in the bottom left two panels of Figure \ref{fig:twobody} we show the self-potential plus the multipole expansion of the external contributions at second and third order (bottom left and bottom center, respectively). Already at second order, where only the effects of the tidal field are added, the potential field exhibits the correct qualitative behavior. At higher orders, the expansion gets closer to the boosted potential, which is by definition the limiting case for infinite expansion terms.

\subsection{The corotating potential}

Finally, in the last panel of Figure \ref{fig:twobody} we show the corotating potential (a.k.a. Jacobi-potential) which is given by adding a centrifugal term to the global potential
\begin{align}
    \phi_{\rm{corotating}} &= \phi - \frac{1}{2} \left| \myvec{\omega} \times \myvec{x} \right|^2  .
\end{align}
\citep[see][for example]{BinneyTremaine2008}. This corresponds to the potential as viewed from a frame corotating with $m_b$ at the orbital frequency $\omega_b$ and where the potential has no time dependence  (for this specific problem). The Jacobi energy $E_J = \frac{1}{2} v^2 + \phi_{\rm{corotating}}(\myvec{x})$ associated with the corotating potential is a strictly conserved quantity if $m_b$ is on a circular orbit around $m_a$. Typically, this frame is used to identify the Lagrange points of the Earth-Sun system. However, it can also be used to find the spatial region where particles can be bound to the mass $m_b$. Particles that do not have enough energy to reach the critical energy level are confined within the critical contour -- also called tidal- or Roche-surface. We note that this region and the two closest Lagrange points, can also be approximately identified in the boosted potential. However, as we shall see, the boosted potential has the advantage that it is also meaningful in more general cases e.g. when $m_b$ is not orbiting $m_a$ on a circle, or when $m_b$ is not on a closed orbit but instead in a free-falling motion. 

\subsection{The tidal hull and effective escape energies}

There are two common definitions used in the literature to identify the tidal boundaries of point masses. One corresponds to the closest saddle point $r_{\rm{saddle}, c}$ in the corotating (Jacobi) potential. Under the distant tide approximation, it can be approximated as
\begin{align}
    r_{t,J} &= R \left(\frac{m}{3M}\right)^{1/3},
\end{align}
and is commonly refered to as the Jacobi-radius \citep{BinneyTremaine2008}. The other commonly used definition is the point where the tidal field from $m_a$ exceeds the self-gravity from $m_b$, which corresponds to the saddle point of the boosted potential $r_{\rm{saddle}, b}$ and becomes
\begin{align}
    r_{t,R} &= R \left(\frac{m}{2M}\right)^{1/3}
\end{align}
in the distant tide approximation -- a.k.a. the Roche-limit \citep[e.g.][]{vandenbosch_2018}. Note that in the case of the Earth-Sun system these approximations are very similar to the actual saddle point radii ($r_{\rm{saddle}, b} = 267.4 R_\oplus$ versus $r_{t,b} = 268.9 R_\oplus$ for the boosted potential, $r_{\rm{saddle}, c} = 232.1 R_\oplus$ versus $r_{t,J} = 234.9 R_\oplus$ for the corotating potential -- where $R_\oplus$ is the Earth radius). 
However, the saddle point notion generalizes more easily to other configurations. Therefore, in later sections we will always refer to the numerically determined saddle point of the boosted potential $r_{\rm{saddle}, b}$ when talking about tidal boundaries.

The difference between the Roche-limit and Jacobi-limit are sometimes considered major problems when modelling what material should get stripped from an orbiting object in analytic approximations \citep[for example]{vandenbosch_2018}. Consider that the difference in tidal radii would correspond to a ratio in tidal volumes of $\frac{3}{2}$ if these volumes would be assumed to be spherical, which can make a big difference in the enclosed mass. The Roche-limit is typically considered a better choice for eccentric orbits, whereas the Jacobi-radius is the optimal choice for circular orbits.

Here we argue that this distinction of different ``spherical'' tidal-radii is an unnecessary complication. As can already be seen in Figure \ref{fig:twobody}, assuming a spherical symmetric boundary of the system at the tidal radius is a very crude approximation. Instead, the energy level relative to the saddle point energy is a better predictor for determining the particles that can leave the system (we demonstrate this in Section \ref{sec:sdisruption}). Now, the saddle point energy in the boosted potential ($\sqrt{2 \Delta \phi_{\rm{esc}, b}} = \SI{11154.6}{\meter\per\second}$) and in the Jacobi potential ($\sqrt{2 \Delta \phi_{\rm{esc}, c}} = \SI{11149.2}{\meter\per\second}$) differ only marginally.
Throughout this article we will exclusively define tidal boundaries and escape velocities from the boosted (not corotating) frame of reference, since this frame can be constructed in any setup and it is therefore much more general. We think that the differences in energy levels are small enough to justify neglecting these subtleties and we are usually not interested in the spatial tidal boundaries at a high precision.

We will argue that \emph{the tidal boundary of a system is best understood as an energy level, not as a radius}.  We will give further quantitative motivation for this perspective in Section \ref{sec:sdisruption}. An obvious criticism might be that energies are not conserved in any system of interest since the potential is explicitly time-dependent. However, we find generally that in the boosted frame, the potential field has only a \emph{weak} explicit time dependence and the energies of bound particles change only weakly over time. Hence, boosted energies can be considered in many contexts as approximately conserved. This implies that energy levels of the boosted potential are quantitatively meaningful and better suited to distinguish between bound and unbound particles than, for example, the energy associated with the self-potential.

\rvtext{However, to also give an estimate of the spatial influence area of an object, we define the 'tidal hull' as the surface of the connected volume inside which $\phi(\myvec{x}) < \phisad$. In practice we approximate it by selecting all particles that are inside this volume and taking their convex hull. We will further sometimes describe the extent of the tidal hull by quoting the distance between the potential minimum and the closest particle outside of the hull as the ``lower tidal radius'' and the distance between potential minimum and farthest particle inside of the hull as ``upper tidal radius''.}

\section{The potential valleys of Dark Matter Haloes} \label{sec:boostedhaloes}
In the last section we argued that a boost operation can turn the potential -- that is dominated by large-scale gradients -- into a locally meaningful potential landscape. In this section we will use this operation to explore the potential valleys of typical haloes and subhaloes in cosmological $N$-body simulations. We will define a novel binding check based on the boosted potential and apply it to define the tidal boundary, the tidally bound mass, and the persistence of haloes. Further, we will use this to distinguish between actual haloes and spurious overdensities in warm dark matter simulations and to show that the tidally bound component of a halo is on average virialized (unlike the component that passes the self-potential binding check).

\subsection{The boosted potential valley}
\begin{figure*}
    \centering
    \includegraphics[width=\textwidth]{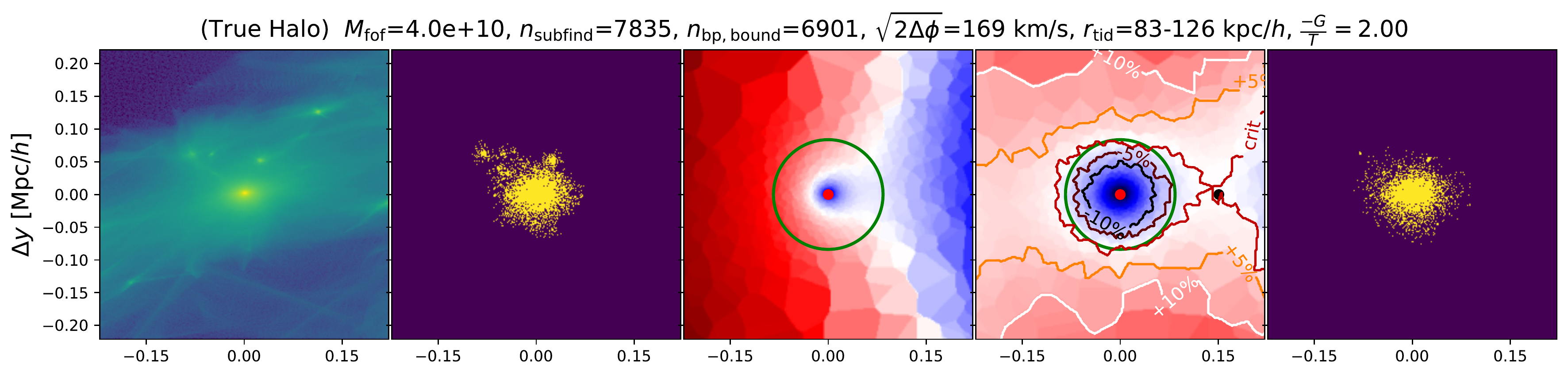}\\
    \vspace{-8pt}
    \includegraphics[width=\textwidth]{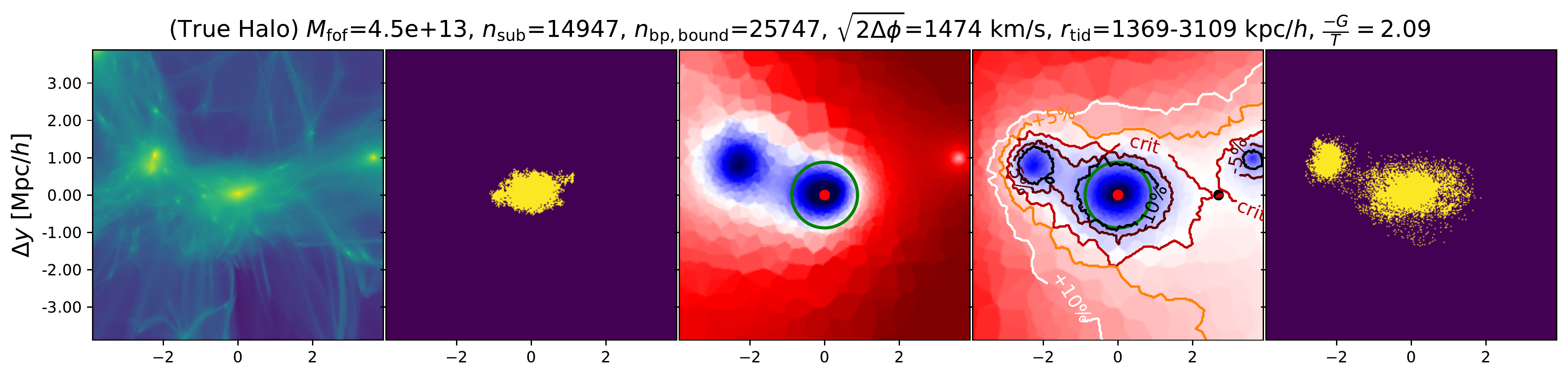}\\
    \vspace{-8pt}
    \includegraphics[width=\textwidth]{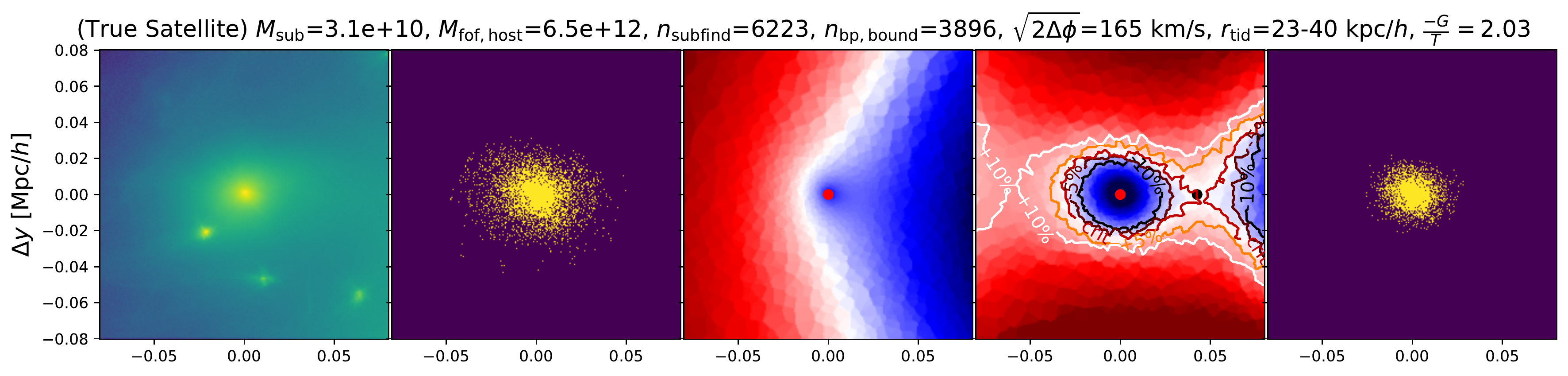}\\
    \vspace{-8pt}
    \includegraphics[width=\textwidth]{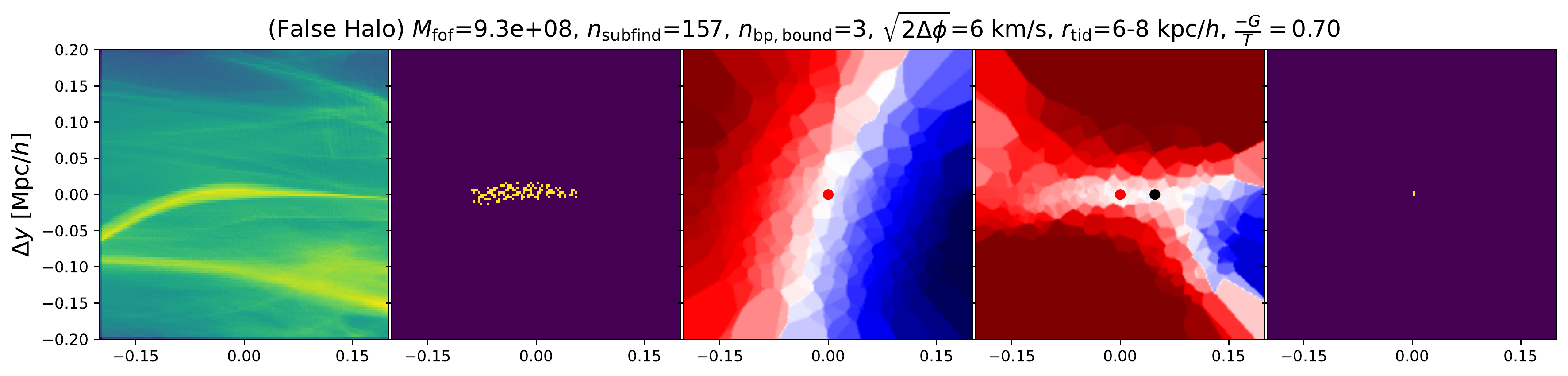}\\
    \vspace{-8pt}
    \includegraphics[width=\textwidth]{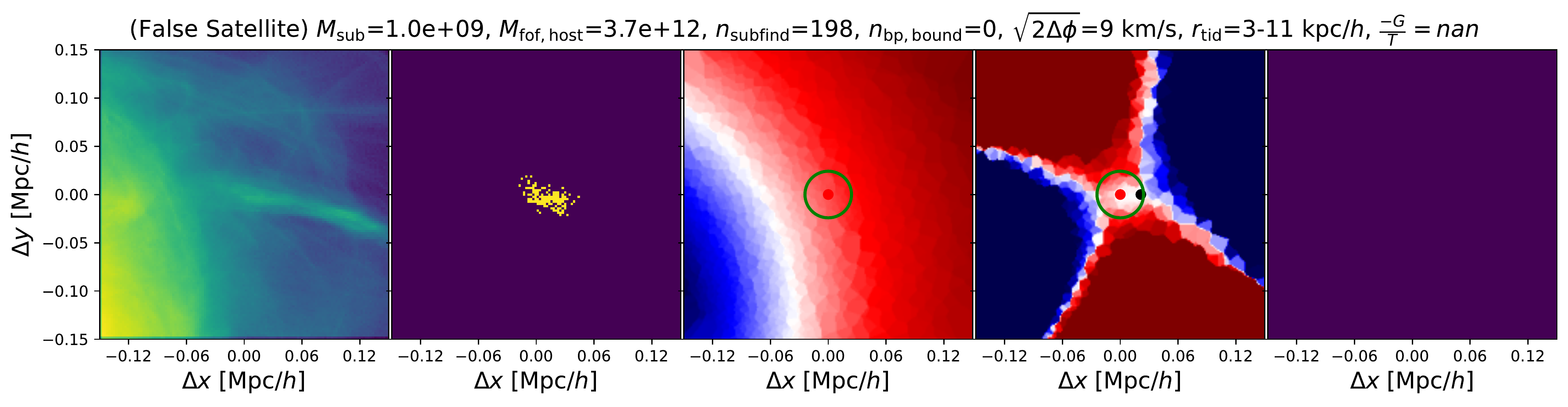}\\
    \vspace{-8pt}
    \caption{Typical examples of how the \subfind{} binding criterion compares to the boosted-potential binding check. The first column shows the sheet-density, around a structure that was identified by \subfind{}, the second column the particles that are bound to the self-potential of that structure according to \subfind{} \rvtext{($n_{\rm{subfind} }$)} and the last panel shows which particles are bound according to the boosted-potential binding check \rvtext{($n_{\rm{bp, bound} }$)}. The third column shows the comoving potential $\phi_c$ and the fourth column shows the boosted potential. \rvtext{Further numbers in the title indicate the halo mass $M_{\rm{fof}}$ / satellite mass $M_{\rm{sub}}$ / host-halo mass $M_{\rm{fof, host}}$ in $M_{\odot /h}$, the effective escape velocity $\sqrt{2 \Delta \phi}$, the lower- and upper tidal radius $r_{\rm{tid}}$ and the virial ratio $-G/T$ (as explained in Section \ref{sec:virialratios})}. The examples from top to bottom are: (1) A small mass CDM halo from the $\SI{20}{\mega \parsec}/h$ simulation. (2) Two larger mass CDM haloes that are shortly before merging from the $\SI{205}{\mega \parsec}/h$ simulation. (3) A CDM subhalo from the $\SI{20}{\mega \parsec}/h$ simulation, orbiting in a 100 times more massive halo. \rvtext{Note that the big object in the center of this image is the subhalo and the central component of the host-halo lies to the right, outside of this image.} (4) A caustic structure from the WDM simulation that appears as a bound halo according to \subfind{}, but exhibits a too low persistence to pass the boosted potential binding check, which properly considers the surrounding tidal field. (5) A similar example, but for a satellite subhalo (according to \subfind{}). }
    \label{fig:boosted_examples}
\end{figure*}

We consider three simulations -- all adopting a Planck-Cosmology \citep{planck_2016} -- to investigate the boosted potential valleys in and around haloes. The first simulation corresponds to a $\Lambda$CDM gravity-only $N$-body simulation with a boxsize of $L = \SI{205}{\mega \parsec}/h$ and with $N=625^3$ particles. This simulation has the same cosmological parameters and realization of the initial dark matter density field as the Illustris TNG 300 simulation \citep{nelson_2019}, and has been run with the cosmological simulation code \textsc{L-Gadget3} \citep{springel_2005,Angulo2020}. We will use exclusively this simulation for quantitative studies in this section. Further, we use a smaller volume $\Lambda$CDM $N$-body simulation with $L = \SI{20}{\mega \parsec}/h$ and $N = 512^3$ particles to provide some images from lower mass cold dark matter (CDM) haloes at increased resolution. Finally, our third simulation corresponds to a warm dark matter (WDM) simulation of a thermal relic with $m_{\rm{x}} = \SI{1}{keV}$ with an effective resolution of $N = 512^3$ particles in a $L = \SI{20}{\mega \parsec}/h$ volume. This WDM simulation has been run with the advanced scheme presented in \citet{stuecker2020complexity} that has been developed to produce reliable WDM simulations without artificial fragments, based on a phase space sheet-simulation scheme \citep{abel2012, shandarin2012, hahn2013, hahn2016, sousbie2016} with an additional switch to N-body dynamics in regions of high phase space complexity. The simulation will be presented among others in more detail in an upcoming publication \citncdm{}. Our main reason to consider the WDM simulation is to show that the boosted potential can be used to solve problems of standard structure finding algorithms, that become especially important in WDM scenarios.

For all simulations we run the \subfind{} algorithm \citep{springel_2001} with a linking length of 0.2   inter-particle separations to detect haloes and subhaloes. We show examples of the density field in the surrounding of typical (sub-)haloes in the left panels of Figure \ref{fig:boosted_examples} and show the particles that are classified as bound according to \subfind{} in the second panel. \subfind{} detects reasonable halo-structures in most cases. However, in the WDM simulation it also detects some elongated overdensities as haloes that would clearly not pass any intuitive notion of ``halo''. As we will discuss in more detail in \citncdm{}, overdensities outside of haloes are expected within dark matter simulations in the form of caustics. They are naturally picked up by the FOF algorithm (which basically detects overdensities), and they are not unbound by the \subfind{} binding check. \subfind{} does not unbind such structures because they appear as bound according to the self-potential. The self-potential neglects the influence of the large surrounding tidal field, which is responsible for preventing such structures from collapsing in all three dimensions.

We calculate the boosted potential as a post-processing step based on the centers suggested by \subfind{}. We note that, in principle, it would be possible to define a structure finder that is based on the boosted potential, completely independently from the framework of FOF algorithms and self-potential binding checks. This is because the critical potential level defines a natural boundary of any halo, and it is therefore not required to impose any other artificial boundary. However, for the purpose of this paper we restrict ourselves to illustrate the differences between binding checks employing the self-potential or the boosted potential, and we leave a more detailed treatment for possible future studies.

For each halo or subhalo identified by \subfind{}, we only use the estimated center $\myvec{x}_0$ and the \subfind{} estimate of the number of simulation particles $n_{\rm{sub}}$ that are part of the structure. Then, we calculate the local mean-acceleration as
\begin{align}
    \myvec{a}_0 &= \sum_{i \in G} \myvec{a}_i \label{eqn:boostedaverage},
\end{align}
where we sum over the $0.8 n_{\rm{sub}}$ particles nearest to $\myvec{x}_0$. These choices are somewhat arbitrary and are mainly motivated by our intuition that the estimate should not depend sensitively on what is happening in the center or the outskirts of the halo. We leave a more rigorous investigation to future studies. We then define the boosted potential valley for each of these objects through the boost operation from equation \eqref{eqn:boostedpot}. 

We display the global and boosted potentials in the third and fourth columns of Figure \ref{fig:boosted_examples}. First of all, we can clearly see how the global potential seems uninformative in almost all cases -- except for very massive haloes -- due to large scale gradients, whereas the boosted potential seems locally meaningful for typical haloes and subhaloes. We can see that we can find a critical energy level where the equipotential lines start opening up on a much larger scale. Haloes appear to be typically well contained inside this boundary. However, there can be differences between the boundary that FOF/\subfind{} draw and the critical contour from the boosted potential. In the second row of Figure \ref{fig:boosted_examples} we can see such an example where two haloes are about to merge. The potential structure suggests that the smaller halo on the left is already bound to the larger halo, which means that it cannot escape its potential well anymore. Note, that the left halo is considered a substructure of the central halo which has a deeper minimum. However, the halo on the right is not considered a part of this structure, since it is connected to a deeper potential minimum which lies outside of this image.

For subhaloes (third row of Figure \ref{fig:boosted_examples}) we see that their associated overdensity typically extends slightly beyond the tidal boundary. Much of this material outside will be lost in the subsequent evolution since it cannot remain bound in such a strong tidal field. We will discuss this in more detail in Section \ref{sec:sdisruption} and focus on the case of haloes in the remainder of this section.

Further, we can see that elongated structures in the WDM simulations (which are not collapsed three-dimensionally) exhibit a potential field dominated by strong external tidal fields. The tidal field is so strong that a well-defined local minimum does not even exist. This is so when the Hessian of the potential has one or more negative eigenvalues. The density (which sets the trace of the Hessian) is not large enough to counter-act the trace-free component that is controlled by the external tidal field. However, such spurious structures can easily pass self-potential binding checks that ignore external contributions through the tidal field.

\subsection{The boosted potential binding check} \label{sec:bindingcheck}
We define the boosted potential binding check (BPBC) which operates in three steps:
\begin{enumerate}
    \item Get $\myvec{a}_0$ by averaging over a group of particles around a candidate center $\myvec{x}_0$ to construct $\phi_{\rm{boost}}$ as described above.
    \item Find the critical contour of $\phi_{\rm{boost}}$ where the valley around $\myvec{x}_0$ merges with a deeper valley. This defines the persistence of a structure.
    \item Consider all particles inside the critical contour as candidate members of the (sub-)halo, but unbind all particles that have enough kinetic energy to escape this energy level.
\end{enumerate}
Note that the tidal boundary can be far beyond the group of particles suggested by \subfind{} and our algorithm works directly with all particles. In Appendix \ref{app:boostednumerics} we provide a more lengthy description of a concrete implementation that uses the \TTK library  \citep{tierny2018} which is based on discrete Morse theory \citep{forman_1998, forman_2002} -- similar to \textsc{DisPerSE} \citep{sousbie_2011}.

We can see immediately that the peculiar structures that do not exhibit a minimum in the boosted potential are unbound by this binding criterion, since they have a persistence of (near) 0. We mark the critical contours of the boosted potential in the fourth column of Figure \ref{fig:boosted_examples} and we have normalized the potential so that the level of the ``escape energy'' is white, potential levels below are blue and levels above are red. In the fifth column of the same Figure we show the particles that are bound according the boosted potential binding check. The structures wrongly identified by \subfind{} appear unbound due to the large tidal fields. We argue that the ignoring of the tidal field is a major shortcoming in the binding checks of all commonly used halo-finding algorithms that use a binding criterion, for example \subfind{} \citep{springel_2001}, \textsc{rockstar} \citep{behroozi_2013}, \textsc{hbt+} \citep{han_2018} and \textsc{subfind-hbt} \citep{springel_2020}.

\subsection{Properties of the boosted valley}
\begin{figure}
    \centering
    \includegraphics[width=\columnwidth]{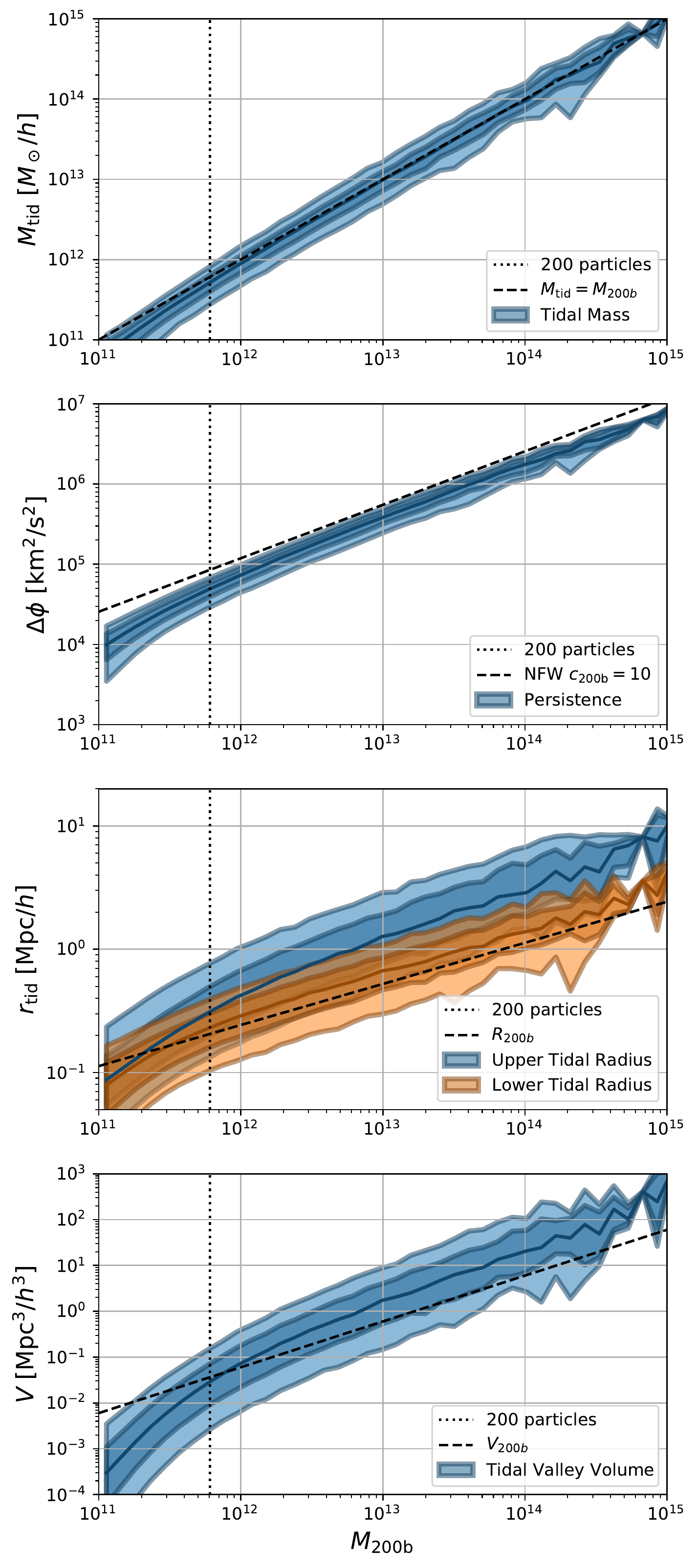}
    \caption{Properties of the boosted valley of $\Lambda$CDM haloes as a function of their mass. The lines and contours show the median and the $65 \%$ and $95\%$ intervals in each bin. The top panel is the tidal mass, the second panel the persistence the third panel the tidal radii and the bottom panel the volume of the tidal valley. All properties appear to be well behaved for haloes with more than 200 particles and they scale as expected (compare slopes of the dashed lines).}
    \label{fig:valley_properties}
\end{figure}

The existence of the critical energy level allows us to define several interesting properties of the potential valleys of haloes. First of all, we can define a new notion of the ``tidal mass'' $\Mtid$ which corresponds to the mass of all particles that pass the boosted potential binding check. We show this in comparison to the mass-definition $M_{200\rm{b}}$ in the top panel of \ref{fig:valley_properties} \rvtext{-- where $M_{200\rm{b}}$ is the mass enclosed in the spherical region which is on average 200 times denser than the mean density of the Universe.}  We can see that typical haloes have $M_{200\rm{b}} \approx \Mtid$, but with a typical scatter of order $50\%$. It seems therefore that this mass notion is well behaved. However, we note that the relation turns down at small masses and we suspect that below 200 particles the tidal mass might be slightly underestimated due to numerical effects. Whether such a notion of the tidal mass is truly useful, will have to be evaluated in future studies that test more rigorously convergence and the importance of numerical parameters. However, we note here that, at least in principle, the tidal mass would be a mass definition that is a priori well justified and does not require any arbitrary density thresholds like common mass-definitions such as \rvtext{the friends-of-friends mass $M_{\rm{FOF}}$ or the masses enclosed in the spherical regions that are 200 times the mean density ($M_{200\rm{b}}$) or 200 times the critical density ($M_{200\rm{c}}$).} The critical energy level is a natural feature of the potential valley and the tidal boundary is a natural limit to the extent of a halo. Therefore, it might be considered as an alternative route for defining halo boundaries and masses in contrast to dynamical approaches based on caustics \citep{shandarin_2021}, phase space folds \citep{Falck_2012} or the splashback radius \citep{diemer_2017}.

Second, we can measure the persistence, or put differently, the effective escape energy for a particle that would reside in the minimum. We show the persistence as a function of $M_{200\rm{c}}$ in the second panel of Figure \ref{fig:valley_properties}. The persistence for an ideal NFW profile \citep{navarro_1997} with vacuum boundary conditions is 
\begin{align}
    \phi_{\rm{NFW}} &= 4\pi G \rho_0 \frac{R_{\rm{200b}}^2}{c_{\rm{200b}}^2} .
\end{align}
which scales as $\phi_{\rm{NFW}} \propto M_{200\rm{b}}^{2/3}$ at fixed concentration. We can clearly see this slope in Figure \ref{fig:valley_properties}, but we note that the normalization of the persistence does not match that of an NFW profile. This is likely because the ideal NFW profile extends until infinity whereas realistic profiles cannot extend much further than the virial radius. It would be interesting to see whether the notion persistence could be used as an alternative notion to mass and whether it might be useful in some scenarios to directly quantify the potential profile of haloes instead of their density profile. In some scenarios like for example gravitational lensing or in the case of the Sunyaev–Zeldovich effect, the potential is arguably more closely related to the observables than the mass or density profiles \citep{lau_2011, vandenbosch_2014, angrick_2015, tchernin_2020}.

\rvtext{In the third and fourth panel we present the spatial extent of the tidal boundary. The third panel shows the lower tidal radius in orange and the upper tidal radius in blue. The fourth panel shows the ``tidal volume'' that is enclosed within the critical contour. These quantities scale all as one would expect -- that is proportional to $R_{\rm{200b}}$ and to $V_{\rm{200b}} = \frac{4 \pi}{3} R_{\rm{200b}}^3$ (indicated through dashed lines). However, it seems that the tidal hull is typically larger than the $R_{\rm{200b}}$ sphere which means that also particles outside of $R_{\rm{200b}}$ can be bound to haloes in principal. However, in practice not many particles might be bound so far out, since the potential field gets very flat close to the critical energy level and therefore only particles within a small energy range are permitted on bound orbits in this outer volume. (Compare for example the first row of Figure \ref{fig:boosted_examples}.)}

It seems that the presented properties of the boosted potential valley are well behaved in general. However, their viability has to be tested more rigorously in a study that investigates their convergence, their dependence on numerical choices and whether they correspond to meaningful boundaries e.g. in halo profiles. An interesting question for such a study would be, whether there is a simple relation between the tidal boundary and the splashback boundary \citep{Falck_2012, diemer_2014, Adhikari_2014, More_2015, diemer_2017}. The splashback boundary, as the outermost region where particles turn around, might roughly correspond to the tidal boundary, which is the outermost region where particles can be dynamically bound.

\subsection{Virial ratios} \label{sec:virialratios}

We have argued that the self-potential binding check considers particles to be bound too easily. Further,  the self-potential binding check depends strongly on the set of particles initially considered, typically through a FOF overdensity criterion. The boosted potential binding check (BPBC) has the advantage that no such preselection is necessary since a natural boundary of potentially-bound particles is given by the critical energy level. We will argue that this leads to the particle populations selected by \subfind{} to be biased to some degree. By biased, we mean that a particle does not always have the same likelihood of being flagged as bound on every point of its orbit. For example, a particle might pass through the FOF preselection when it is at the pericenter of its orbit, but it might be discarded when it is at the apocenter (which might lie outside of the FOF-Group). 

We will now see that this might have an effect on the virial ratio measured for halos in cosmological simulations. We define the virial as
\begin{align}
    G &= \langle \myvec{a} \cdot ( \myvec{x} - \langle \myvec{x} \rangle ) \rangle,
\end{align}
where, in principle, $\langle \cdots \rangle$ denote time averages, but under the assumption of ergodicity they can be replaced by an ensemble average over a proposed group of particles. For this group, we use either the particles that are bound according to \subfind{} or according to the BPBC. The virial theorem states that for a steady system the following relation should hold:
\begin{align}
    - \frac{G}{T} &= 2 \label{eqn:virialtheorem}
\end{align}
where we call $-G/T$ the virial ratio and $T$ is the mean kinetic energy
\begin{align}
    T &= \frac{1}{2} \langle (\myvec{v} - \langle v \rangle )^2 \rangle
\end{align}
We remind the reader that \eqref{eqn:virialtheorem} is the most fundamental version of the virial theorem. The more commonly used ``potential virial theorem'' depends on some additional assumptions, such as the non-existence of external contributions to the potential, and it is thus not obvious what is the correct potential definition to use when external contributions exist. In contrast, the force virial theorem should hold straight-forwardly given the global acceleration field $\myvec{a}$.

\begin{figure}
    \centering
    \includegraphics[width=\columnwidth]{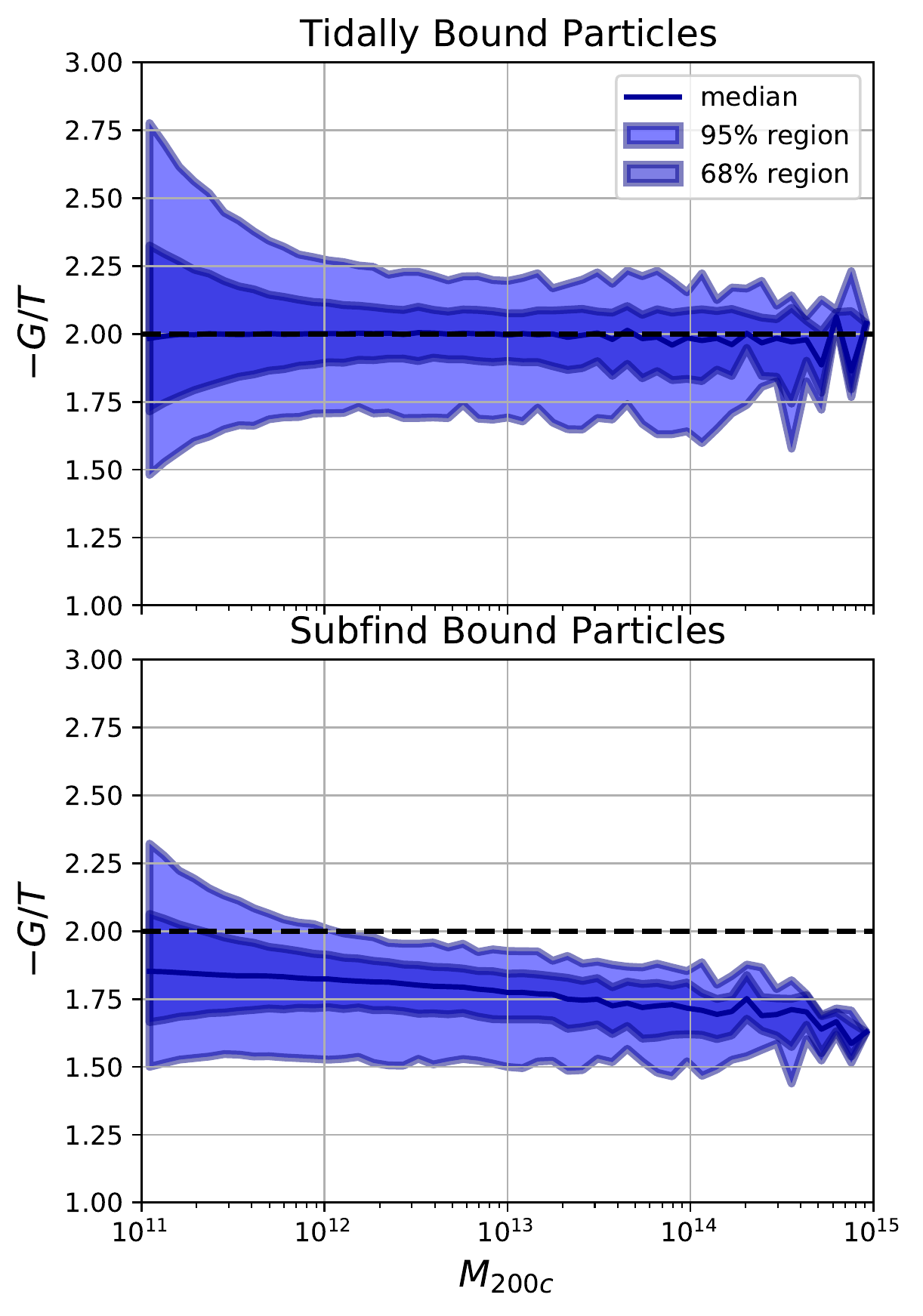}
    \caption{The Virial ratio of haloes from a CDM N-body simulation when defined over the bound populations according to the boosted potential binding check (top) or the \subfind{} bound particles (bottom). The tidally bound particles seem to be well virialized on average whereas the \subfind{} bound population systematically deviates from the expected virial ratio of 2. We think that the virial ratios from \subfind{} are biased, because it does not select particles with the same likelihood at each point of their orbit.}
    \label{fig:virialization_versus_mass}
\end{figure}

We show the virial ratios we obtain when averaging over particles that are bound according to the boosted potential in the top panel of Figure \ref{fig:virialization_versus_mass} and in the bottom panel for the particles that are bound according to \subfind{}. The result is quite striking: the BPBC selected sub-system is clearly virialized on average, centering on the expected value of two, independent of halo mass. On the other hand, the \subfind{} selected particles show virial ratios that are biased low, with a significant dependence on mass. This systematic offset in the virial ratios has already been noted before -- see for example the measurements from the Millennium simulation in Figure 4 of \citep{Bett2007}. Some authors have argued that these ratios can be corrected by including the effect of surface terms in the virial theorem \citep{shaw_2006, Davis2011}, but, even then, the virial ratios do not seem to align optimally \citep{Davis2011}. 

Our results suggest that the virial ratio typically lying below 2 means that \subfind{} prefers to select particles at that point of their orbit where their kinetic energy is higher (the pericenter) and is less likely to include the same particles when their kinetic energy is lower (at their apocenter). If this were the case, it would not be justified to assume ergodicity of the selected subsystem and the time-averages in the virial theorem cannot be assumed to be equivalent to instantaneous averages over this (biased) population. 

We propose this to be investigated in a quantitative study where the orbits of individual particles are followed explicitly and it is checked whether the likelihood of detection varies along the orbit. We note that, in principle, the virial theorem should hold for time-averages of orbits of individual particles and it would be interesting to see whether the instantaneous FOF preselection acts effectively as a time-dependent weighting function.

\subsection{Discussion}

We have shown that the boosted potential binding check has, in principle, a variety of advantages when compared to traditional self-potential binding checks: (1) It does not require a FOF preselection. (2) It is based on an actual critical feature in the potential field and does not require arbitrary thresholds. (3) It gives natural notions of the tidal-boundary of haloes and explains why haloes do not extend until infinity. (4) It produces the expected virial ratio of two. (5) It can distinguish between actual haloes and caustic overdensities and (6) the used notion of boundedness can naturally be generalized to other cosmic-web structures as we discuss in the next section.

That said, we have not presented an optimal algorithm that would take advantage of all of these benefits. A future study could address (1) whether there exists a viable and simple way to come up with proposed halo centers that does not require the outputs of other structure finders. As an example, an algorithm could seed centers at each region where all eigenvalues of the Hessian of the potential are positive. (2) Whether there exist better alternatives to define the acceleration of the boosted system. Ideally, the accelerations could be inferred from topological/geometrical features of the potential field. We will discuss this possibility in Section \ref{sec:advancedmath}. (3) Whether such an improved algorithm shows good convergence behavior and (4) whether the particles that are considered as unbound are actually leaving the system under consideration. We think that there exists sufficient motivation for such a follow-up study. However, for the remainder of this paper, we will explore other application areas of the boosted potential.

Beyond these numerical questions we think that there might exist physical motivation to use the boosted potential to investigate the tidal boundaries of haloes more rigorously in future studies. Such studies might help to understand the relation between large scale tidal fields and the alignment of haloes \citep{stucker_2021}, to determine what is the most natural way of truncating NFW haloes \citep{drakos_2017} and answer whether the dynamical boundary of haloes -- often called the splash back radius \citep{diemer_2014}  -- coincides with the tidal boundary.

\section{The boosted potential and the cosmic web} \label{sec:cosmicweb}
\begin{figure*}
    \centering
    \includegraphics[width=0.9\textwidth]{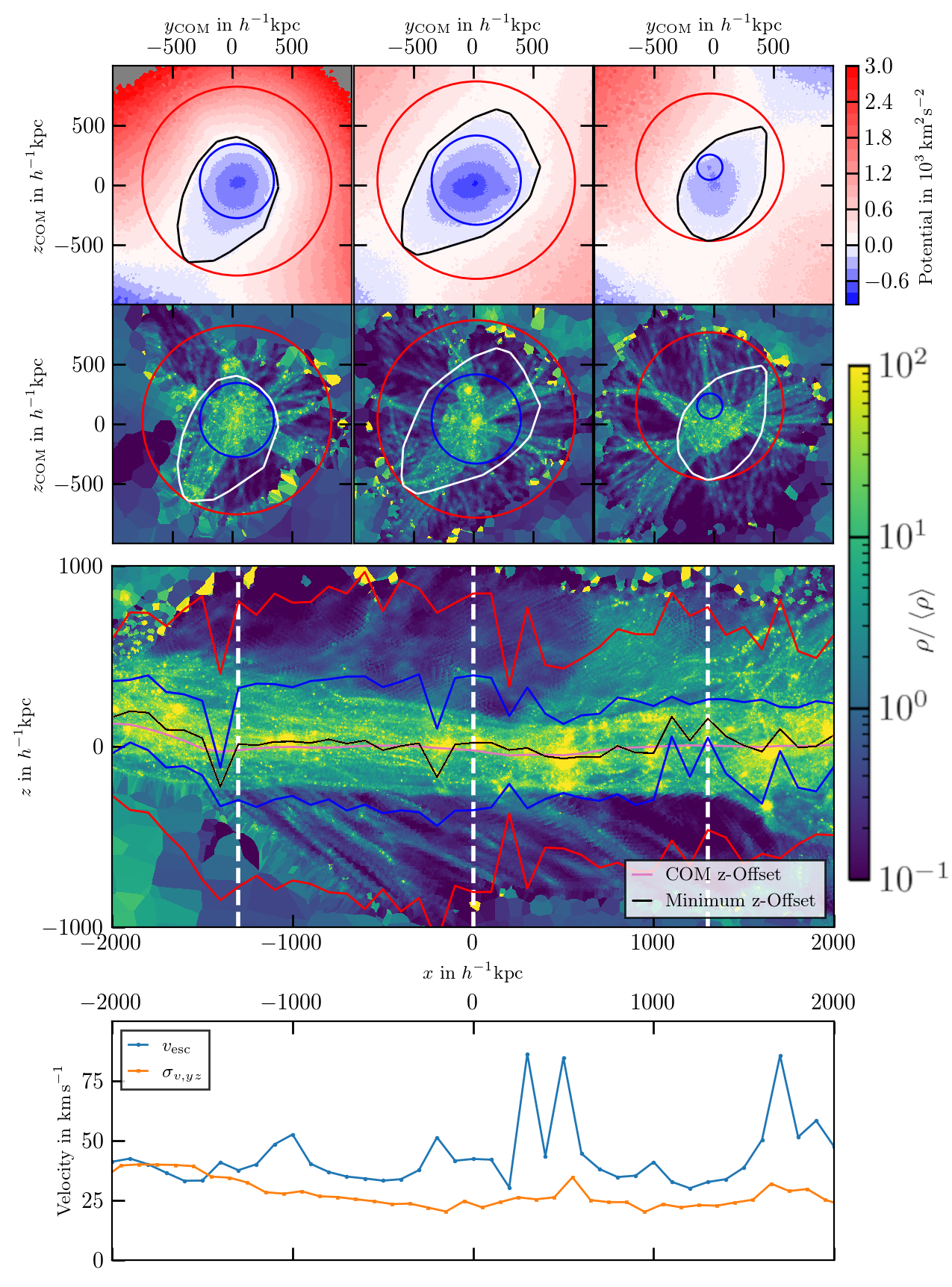}
    \caption{The boosted potential in a short, rather isolated filament extracted from the Illustris-TNG slices. Blue is the lower and red the upper tidal radius. \emph{Top:} Three slices across the filament, showing both the boosted potential and the density for each. The black/white contour follows the critical line at $\phisad$. Center: A slice showing the density along the filament axis and with the positions of the slices in the upper panel indicated. Bottom: The $y,z$-escape velocity at each $x$ location and the corresponding $y,z$-velocity dispersion. We see that (up to numerical issues) it is possible to define a meaningful tidal boundary of the filament and that the particles seem indeed to be bound to the corresponding energy level. \rvcom{(This Figure has been changed)}}
    \label{fig:fil_slices}
\end{figure*}
In this section we motivate that, in principle, the language of the boosted potential can be applied to the cosmic web as well and that the suggested notion of bindedness is also meaningful for other cosmic structures, like filaments and walls -- for which a self-potential binding criterion is not feasible due to the dominance of external tides.

\subsection{The potential landscape of the cosmic web}

It is instructive to start this section with a thought experiment: Imagine we had a potential landscape that was constant with time and we had a large set of massless test particles orbiting in this landscape. We could define structures in this landscape as spatial regions that different particles can be confined to. Since energies are conserved, it would be straight forward to identify structures of different types in this landscape. The first kind of structure is given by potential contours that are closing in all three dimensions around some minimum. We have seen many examples of such contours in the last section and we could roughly identify these structures with haloes. A particle that has an energy level below the critical ``escape energy'' would be confined to the region below the corresponding energy level. At the critical energy level the contours open up in one dimension, but (typically) remain closed in the other two dimensions. We could define a new type of structure, a filament, where particles can move freely along this one dimension, but are still confined in the other two directions. Further, we could define wall-like structures by level sets that are open in two dimensions and closed in one, and void structures where the level sets are open in all dimensions\footnote{We mean open and closed here always in relation to the descending direction.}. These structures could be nested, for example, if we consider a halo that lies in a filament, all particles bound to the halo would also be confined to the region of a filament the halo is part of, but some particles are just bound to the filament and could move in and out of the embedded halo.  All these structures would be well defined and could be disentangled with the tools of Morse Theory -- see for example  \citet{sousbie_2011} and \citet{sousbie_2011b} for a rigorous treatment in the case of the density field of cosmological simulations.

Now, imagine, we would distort the same potential landscape. We would add different large scale potential gradients to each structure and we would move each potential structure in time according to the associated acceleration (and velocity). The potential is explicitly time-dependent now, and it is very complicated to understand which particles are confined to which regions, since the level sets lost their meaning as they got distorted by the large-scale gradients. We think that this illustrates the situation we face with the cosmological potential field. Many of the intuitions about restricted regions in phase space should still hold (approximately), but we are missing the necessary tools to identify them in this complicate distorted landscape.

One might argue here that the potential field is therefore the wrong field to consider and instead structures should be identified based on the density field \citep{sousbie_2011, sousbie_2011b, busch_2020}, the tidal field \citep{hahn_2007}, the velocity field \citep{hahn_2015, buehlmann_2019} or Lagrangian criteria \citep{shandarin_2012, Falck_2012, neyrinck_2015, stuecker2020complexity}. While all of these are viable methods for finding the skeleton of the cosmic web, none of them can employ physical binding criteria and predict what particles remain confined to a structure, and in most cases it is not clear what is a natural boundary of structures. We would define a ``structure'' as a (possibly moving) sub-region of space that particles can be confined to. If this notion of structure is adopted, one has to conclude that most of the above methods detect structures by their \emph{features}, but not directly by their \emph{definition}. However, we want to show here that such confined regions are, in principle, encoded in the potential field.

Since we do not have the mathematical tools to rigorously disentangle this complicated potential field as a whole, we will instead show with a single filament as an example, that its potential landscape becomes meaningful when investigated from a boosted frame of reference.

\subsection{The potential of a filament}

We investigate the dynamics in a single short ($\sim\SI{6}{\Mpch})$ filament in a cosmological context as shown in \autoref{fig:fil_slices}. This filament was extracted from the Illustris-TNG 100 simulation and re-simulated at much higher resolution ($m_\mathrm{part,DM}\approx 10^3\si{\msol\per\h}$) as a zoom simulation \citep{busch2021}.

We expect a filament to be quite similar to a halo regarding the dynamics in the (transverse) directions orthogonal to its axis. Shell crossing has occurred in these dimensions and matter is bound to the local structure of the filament. However, the longitudinal dimension had no shell crossing and material can descend towards its ends, where the potential is generally lower.

The absence of an easily accessible reference structure, such as the density-identified haloes above, and the presence of haloes as substructure make it harder to find an appropriate boosted system for the filament. By visual experimentation, we define a coordinate system where the filament is roughly aligned with the x-axis. We then investigate the potential field in slices orthogonal to the x-axis. To define the boosted potential in every slice, we average the potential gradient in a $\SI{300}{\kpch}$ transverse disk through the filament centred on the centre of mass calculated in a $\SI{500}{\kpch}$ disk around the origin. 

We show the resulting boosted potential slices at three different locations in the top row of \autoref{fig:fil_slices} using a coordinate system again centred on the COM in each slice as defined above. With the \TTK library we determine the saddle point in each slice and set the corresponding energy level to 0 so that tidally bound regions correspond to negative potential values. When there are several minima--saddle point pairs, we select the pair with the highest persistence out of those that have the minimum in the region that we used to calculate the boost (with radius $\SI{300}{\kpch}$). We mark the convex hull of the points on the critical energy level by black contours and mark the lower and upper tidal radius with blue and red circles around the minimum, respectively.

First of all, we note that in each of the slices, we can find a minimum and closing surrounding contours in the boosted potential. We have also looked at the global potential in the same slices, and it is completely dominated by a uniform gradient term and does not show any local minimum. Further, we note that we can identify the critical contour in each slice and a corresponding escape energy level. The upper and lower tidal radius are a bit less well defined, since the minimum can jump to substructures inside different slices. A halo inside the filament will typically have a lower potential minimum than the overall filament. However, the upper and lower tidal radius are just crude notions of the extent of the filament, and the main reason we defined these is so that we can easily show an estimate of the extent in the projections along the filament.

\subsection{The tidal- and density boundaries of the filament}

The second row in \autoref{fig:fil_slices} compares the critical contour and tidal radii with the dark matter density in the same slices. We see that the overdense region that is associated with the filament is well contained within the critical contour in all slices. The region within the contour is not filled out completely until its boundary. We can see this behavior for haloes and subhaloes as well (compare Sections \ref{sec:boostedhaloes} and \ref{sec:sdisruption}) and this is likely related to the fact that the potential gets very flat close to the boundary. However, the smaller axis of the contour delineates relatively well the boundary of the filament and therefore the transverse extend of the filament is typically reasonably well approximated by the lower tidal radius -- except for the pathological cases where the minimum jumps to a halo, like in the third slice of Figure \ref{fig:fil_slices}.

In the third row of \ref{fig:fil_slices} we compare the dark matter density in a slice along the filament to the upper and lower tidal radius in red and blue and the projected COM and potential minimum in pink in orange. Here we defined these quantities through a large number of different slices that were constructed in the same way as the three examples in the top panels of Figure \ref{fig:fil_slices}.

Just as in the slice plots above, we find that the lower tidal radius gives a reasonable estimate of the extent of the dark matter splashback around the filament axis. Over wide ranges the blue curves trace the density jump around the filament more or less correctly. For $x>0$ the connection is tighter, even when we misidentify the potential minimum in a few cases, and hence COM and minimum have a very different offset. However, one should not overinterpret the correspondence here, since these tidal radii are just estimates of the extent of the critical contour. We can only say for sure that this critical contour should always lie between the two radii in this projection. 
\subsection{Binding state along the filament}

In the bottom panel of \autoref{fig:fil_slices} we show the transverse escape velocity $v_\mathrm{esc}=\sqrt{2(\phisad-\phi_\mathrm{min})}$ from the boosted potential minimum of each slice.  The escape velocity is rather constant along the filament with most values falling into the range \SIrange{30}{45}{\km\per\second} and only a handful of clear peaks exceeding this range. A look at the minimum offset and density field in the longitudinal density panel of \autoref{fig:fil_slices} reveals that these peaks are associated with haloes. The escape velocity from the minimum of these is of course larger due to the additional potential difference between the halo centre and the filament centre.

Further, we show in the same Figure the transverse velocity dispersion $\sigma_{v,yz}$ as calculated in \SI{100}{\kpch} thick slices across the filament with a radius of \SI{500}{\kpch}. In each slice we determined the dispersion of the velocity in the $y$-$z$ plane of the high resolution dark matter particles. The choice of radius ensured that the whole filament within the splashback boundaries was included and dominating the numbers. We can see that typical transverse velocities are well below the escape velocity of the filament -- like one would expect for a bound structure. However, at $x < \SI{-1500}{\kpch}$ this does not seem to be the case anymore. We speculate that this is so, since the filament is not so well aligned with the $x$-axis in that regime and the longitudinal and transverse directions get mixed. A more sophisticated technique for tracking the filament would probably solve this.

\subsection{Discussion}

In this section we have shown, with the example of one filament from a cosmological simulation, that, in principle, the boosted potential can also be used to understand the potential landscapes of structures other than haloes. From a boosted frame of reference, the potential contours of a filament close in the dimensions orthogonal to its axis and they open along its axis. It is possible to define the tidal boundary and escape velocities in orthogonal slices and we have seen that these appear meaningful by comparing them to the density structure and the velocity structure of the filament.

Our main motivation here was to show that the structure of haloes, filaments and walls is well encoded in the potential field and that, in principle, they can all be understood in the same language. However, we are lacking good methods for disentangling the global potential field due to the omnipresent large-scale gradients. The boosted potential opens us a window to create locally meaningful extracts of the potential landscape. However, we would like to encourage other researchers to search for more globally applicable tools. Just as Morse Theory would arguably be the perfect tool to disentangle the potential landscape of a static potential, we imagine that a generalized version of Morse Theory might be able to disentangle the complicated time-dependent large-scale-gradient-dominated potential landscapes that we face in cosmological simulations.

\section{The ``Deforming Bowl'' Picture of Tidal mass loss} \label{sec:sdisruption}

We have seen that the boosted potential can be useful to physically understand the boundaries of halos and of filaments. In this section we show that it can also be useful to understand the tidal stripping that occurs in subhaloes that fall into a larger dark matter halo. The main point that we want to make is that the tidal boundary of a subhalo is best understood as an energy level in the energy associated with the boosted potential. Consideration of the boosted potential leads us naturally to a qualitatively and quantitatively powerful picture of tidal mass loss: the ``deforming bowl'' picture. In this picture, the boosted potential valley corresponds to a bowl which holds water (representing particles). The boundary of the bowl is deformed over time through the tidal field -- letting the water that lies above the saddle point level of the deformed boundary escape. We illustrate this qualitatively in Figure \ref{fig:deformingbowl}, but a more detailed explanation follows:

\begin{figure}
    \centering
    \includegraphics[width=\columnwidth]{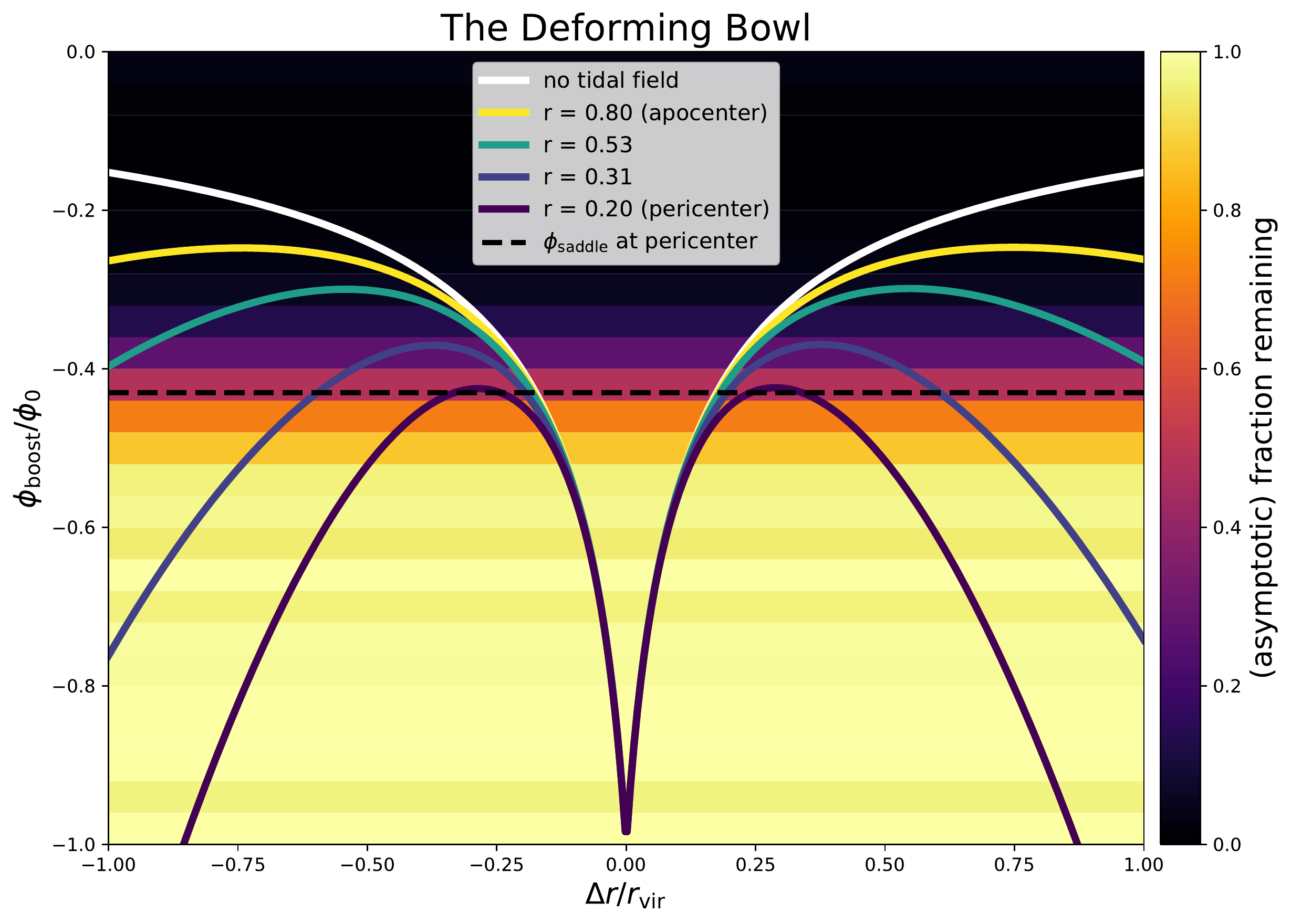}
    \caption{Illustration of the ``deforming bowl'' picture of tidal mass loss. The plot shows the boosted potential in the radial direction, where the tidal field lowers potential levels. Different lines show it at different times where the subhalo is at different radii and therefore exposed to tidal fields of different strength. The horizontal colours in the background show the fraction of particles that is remaining in different energy bins after many orbits ($\gtrsim 10$). The potential landscape of the subhalo is permanently deforming along the orbit with the variation of the tidal field. However, the asymptotically remaining particles are mostly those particles which do not have enough energy to pass through the saddle point at pericenter.}
    \label{fig:deformingbowl}
\end{figure}

The intuitive explanation of tidal mass loss that is typically presented in the literature is well-summarized in the word ``tidal-stripping''. The idea is that the tidal field shaves off material of the satellite subhalo that passes beyond the tidal radius. We note that this qualitative explanation has inspired the choices that have been made in numerous quantitative descriptions, e.g. where the mass loss rate is modelled to be proportional to the amount of mass outside of the instantaneous tidal radius \citep[for example]{Taylor_2001, vandenbosch_2018, errani2021}. However, this picture neglects important aspects of the dynamics: (1) The tidal boundary is by no means spherically symmetric (2) The tidal field also modifies the orbits and energy levels of particles that are inside of the tidal radius. (3) Quantitative results depend strongly on the adopted definition of tidal radius \citep{vandenbosch_2018, errani2021}.

Here, we propose the ``deforming bowl'' picture as a simple and intuitive alternative perspective where the tidal mass loss can be rather understood as a ``tidal overflow''. We argue that this is the natural way of understanding the problem from the perspective of energy space, and that many of the findings of \citet{errani2021} regarding the asymptotic remnants of subhaloes emerge naturally.

\begin{figure}
    \centering
    \includegraphics[width=\columnwidth]{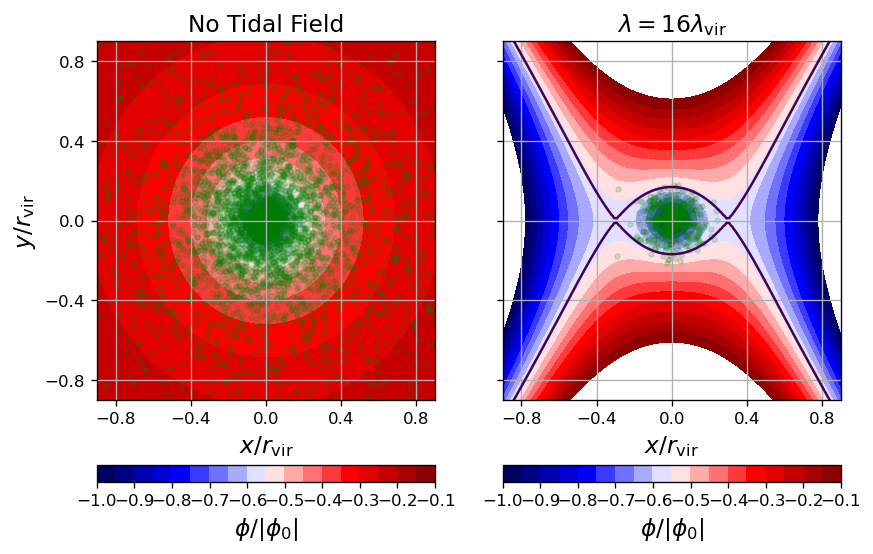}
    \caption{Potential field of the tidal experiments that we conduct in Sections \ref{sec:adiabatictid} - \ref{sec:adiabaticescape}. Left: the NFW potential and a realization of particles orbiting in the potential. Right: The potential landscape and the remaining particles after applying a tidal field. The tidal field has lowered the escape energy level (coloured as white contour and marked as a black line) so that most particles with higher energies have left the system.}
    \label{fig:tidalexperiments}
\end{figure}

We can think of the potential valley of our subhalo as a bowl which can hold water up to a certain (energy-)level. If there is no tidal field, this maximal energy level is $0$ whereas the minimum of our bowl lies at some negative energy level $\phi_{\rm{min}}$. E.g. $\phi_{\rm{min}} = -\phi_0$ with
\begin{align}
    \phi_0 &= 4 \pi G \rho_c r_s^2 \label{eqn:nfwphi0}
\end{align}
in the case of the NFW profile. This situation is displayed in one dimension as a white line in Figure \ref{fig:deformingbowl} and in two dimensions in the left panel of Figure \ref{fig:tidalexperiments}.

Now, if a tidal field is applied, it modifies the curvature of our bowl. In the directions associated with negative eigenvalues of the tidal tensor, the bowl gets additionally bent upwards. In the directions that are associated with positive eigenvalues it gets bent downwards -- creating a saddle point and lowering the minimal energy that is needed to reach the boundary. This is illustrated in the radial direction, which is associated with the ``downwards bend'', in Figure \ref{fig:deformingbowl} and in the full potential field in the right panel of Figure \ref{fig:tidalexperiments}. If we were to put any particle beyond the saddle point, it would get accelerated away from our subhalo. Further, particles which are close to the center, but have high enough energy to leave the bowl, will also ultimately exit the system. In the ``deforming bowl'' picture we can imagine that the tidal field is pushing down the rim of our bowl, allowing to escape all water that lies on a higher level than the new boundary -- thereby adapting the water level to the height of the new boundary -- indicated as black contour in Figure \ref{fig:tidalexperiments}.

Therefore, in the ``deforming bowl'' view, if we apply a tidal field slowly, and further assume that energies are approximately conserved, it will only depend on the energy which particles escape and the only information that is needed about the tidal field to predict this is the saddle point energy. This saddle point energy will typically only depend on the largest eigenvalue of the tidal tensor -- the one bending down the bowl the most. We will explore this quantitatively in Sections \ref{sec:adiabatictid} - \ref{sec:adiabaticescape} and show that this energy budget is a good predictor of which particles escape a system with a slowly changing tidal field.

However, in realistic scenarios a subhalo is orbiting in the potential of a larger halo. The tidal field continuously changes, typically oscillating between its largest amplitude at pericenter and its lowest amplitude at apocenter. If this oscillation was very slow in comparison to the dynamical time inside the subhalo, we would predict that all mass that exceeds the saddle point energy at pericenter will be lost in the first pericenter passage and subsequently the mass would be constant, since the escape energy level will never again get lower than the water level. However, in practice the tidal field will often change much faster than the dynamical time inside the subhalo. Therefore, not all particles that have high enough energy can escape in their first pericenter passage. (The escape route is only opened for a short time). However, if the subhalo passes the pericenter many times, we'd expect that all particles with high enough energy will eventually leave the system. This can be seen in Figure \ref{fig:deformingbowl} where we show the fraction of particles that is remaining for a subhalo after a large number of orbits ($\gtrsim 10$) between a pericenter of $0.2$ host virial radii and an apocenter of $0.8$ host virial radii. We will describe the corresponding experiments in more detail in Section \ref{sec:orbitsims}.

Therefore, the ``deforming bowl'' view suggests that which particles escape the system in the long run depends only on the largest eigenvalue of the tidal tensor at pericenter. Note that this qualitative picture is in excellent agreement with many numerical findings of \citet{errani2021}. They have found that the bound mass and the density profiles of asymptotic tidal remnants -- that is the remainders of subhaloes after they have gone through a large number of orbits (5-30) -- depends mostly on the initial profile of the subhalo and the tidal field at pericenter \citep[see also][]{nadler_2018}, and is almost independent of other orbital parameters (e.g. the eccentricity). Other orbital parameters control mostly for the rate at which the mass is lost, since they control how long a subhalo spends near pericenter -- or, in our picture, how long the escape point of the deformed bowl is opened.

We note that energy-based views of tidal stripping have already been proposed in other studies. E.g. \citet{aguilar_1985, aguilar_1986} have estimated the escape fractions of tidally interacting spherical galaxies by the fraction of particles which increase their binding-energy above the escape energy in the impulse approximation.  \citet{choi_2009} have argued that mass loss can be understood as an inside-out process in energy space. Based on the same recognition, \citet{drakos_2017} developed a procedure to create phase space distributions of tidally truncated NFW profiles, by removing all particles beyond some radius and by iterative unbinding of particles with energies beyond the vacuum escape energy. \citet{drakos_2017} show that the density and phase space profiles of stripped subhaloes are well approximated by a one-parameter model that only requires knowledge of the cutoff energy. In \citet{drakos_2020} the same authors have combined this approach with a mass loss criterion to predict the mass evolution of objects. We agree with the energy-space perspective taken by these authors and we would like to present further arguments for energy-truncation based models. 

The goal of this section is to test whether the conclusions that our proposed ``deforming bowl'' view suggests hold quantitatively. Therefore, we will make a small set of idealized simulations. All of these simulations use a set of massless particles that orbit in analytic potentials. In the first set of simulations (Sections \ref{sec:adiabatictid} - \ref{sec:adiabaticescape}), we will use an NFW potential as the potential landscape and a very slowly increasing tidal field to infer the adiabatic limit of tidal stripping. In the second set of simulations (Section \ref{sec:orbitsims}), the potential will be that of an NFW profile of a subhalo that is orbiting in the NFW potential of a host-halo. We want to point out clearly that these simulations do not resemble realistic scenarios, since when a halo loses mass, its mass- and potential-profile will change. Subsequently, more particles might escape since their energy levels might have changed due to this response of the halo-structure. These effects are important for quantitative predictions for realistic halos, but are explicitly excluded from the simulations in this section. However, we will show in a sub-sequent paper \citsubhalo{} that the same qualitative picture still holds in simulations with full self-gravity and can be used for making simple, but accurate, predictions for the asymptotic mass loss of orbiting subhaloes.

\subsection{Tidal overflow in the adiabatic limit} \label{sec:adiabatictid}

We set up a set of simulations to test which particles will leave an NFW halo in the presence of a tidal field. For these simulations we assume a potential
\begin{align}
    \phi(\myvec{x}, t) &= \phi_{\rm{NFW}}(\norm{\myvec{x}}) - \frac{1}{2} \myvec{x}^T \Tid(t)  \myvec{x} \label{eqn:tidpot}
\end{align}
where we grow the tidal field slowly over time
\begin{align}
\Tid(t)=
\begin{cases}
0  &\text{if } t \leq 0,\\
\frac{t}{\tau} \Tid_f &\text{if } 0 < t \leq \tau\\
\Tid_f &\text{if } t > \tau
\end{cases}
\end{align}
 up to some maximum value $\Tid_f$ at time $\tau$. The potential $\phi_{\rm{NFW}}$ corresponds to the potential of an NFW halo with $c=10$
 \begin{align}
    \phi_{\rm{NFW}}(r) &= - \phi_0 \frac{r_s}{r} \log \left( 1 + \frac{r}{r_s} \right)
 \end{align}
 with $\phi_0$ from equation \eqref{eqn:nfwphi0}. Note that we keep the potential $\phi_{\rm{NFW}}$ fixed, independently of what particles have escaped the potential well. This is an important simplifying assumption, since normally the halo's potential field would change after particles are removed. However, here we are just interested in seeing how well we predict the particles that leave the system if the self-potential is kept fixed. For sure those particles will also leave the system if the self-potential gets reduced due to the escaping mass, but in that case additional particles would also escape.

We follow the method described by \citet{Errani2020} to create a particle realization of an NFW halo with $c=20$ where we sample the energies directly from the numerical solution to Eddington's inversion formula \citep{Eddington1916} up to a radius of $8 r_{200,c}$ where  $r_{200,c}$ is the radius inside which the halo (with mass $M_{200,c} = 10^6 M_\odot /h$) has 200 times the critical density of the universe. Note that we set the halo up to such large radii that the end-result of our simulations only depends on the tidal truncation and not at all on the artificial truncation that we imposed when creating the initial conditions. We let the particles orbit in the potential defined through \eqref{eqn:tidpot} where we choose for the tidal tensor
\begin{align}
    \Tid_f &= \begin{pmatrix}
\lambda & 0 & 0\\
0 & -0.3\lambda & 0 \\
0 & 0 & -0.7 \lambda
\end{pmatrix}
\end{align}
We note here that only the value of the largest eigenvalue ($\lambda$) has relevant quantitative impact on the results and that we choose the other components to ensure a trace-free tidal field and to avoid unlikely symmetries. \rvtext{We run each simulation with 200000 particles and choose the time-steps so that the dynamical time at $0.1 r_s$ is still resolved with 20 time-steps. Both these choices are much more precise than what is required to achieve converged results on the presented figures.}

We create several simulations with different values of $\lambda$. We show an example of such a simulation in Figure \ref{fig:tidalexperiments}. \rvtext{Note that the green points in this figure correspond to all particles, not just bound ones. However, most unbound particles have already moved far away from the subhalo in the right panel.} We usually phrase our results in terms of the tidal field that is needed to create a saddle point in the potential of an NFW halo at the virial radius $r_{200,c}$:
\begin{align}
    \lambda_{\rm{vir}} &= \frac{- \partial_r \phi_{\rm{NFW}}(r_{200,c})}{r_{200,c}}
\end{align}
The results of these simulations depend very slightly on the timescale $\tau$ that is imposed. However, in the limit of large $\tau$ the dependence disappears completely and we approach the adiabatic limit. We choose $\tau = 20 T_{\rm{vir}}$ and integrate the simulation further until $t = 40 T_{\rm{vir}}$ (at a constant tidal field) where 
\begin{align}
    T_{\rm{vir}} = \frac{2 \pi r_{200,c}}{v_{\rm{circ}} (r_{200,c})}
\end{align}
is the circular orbit time at the virial radius. With this choice we are always well in the adiabatic regime. However, we show in Figure \ref{fig:energyadiabticlimit} how much the energies change due to the tidal field for different values of $\tau$. Energies defined through the self-potential oscillate whereas those of the boosted potential seem to converge to values that are almost unchanged in the adiabatic limit. We think that the boosted energy is almost constant in the adiabatic limit, since a trace-free tidal field approximately lowers the energy level of an (on-average) spherical orbit as much as it increases it. In this case, the boosted potential seems a much more natural energy definition than the self-potential. However, independently of the energy-definition, the change in energy is very small in comparison to the persistence of the potential $\phi_0$. We can easily neglect it in the qualitative picture that we presented previously and it seems irrelevant for quantitative predictions.

\subsection{Predictors of escaping particles} \label{sec:adiabaticescape}

\begin{figure}
    \centering
    \includegraphics[width=\columnwidth]{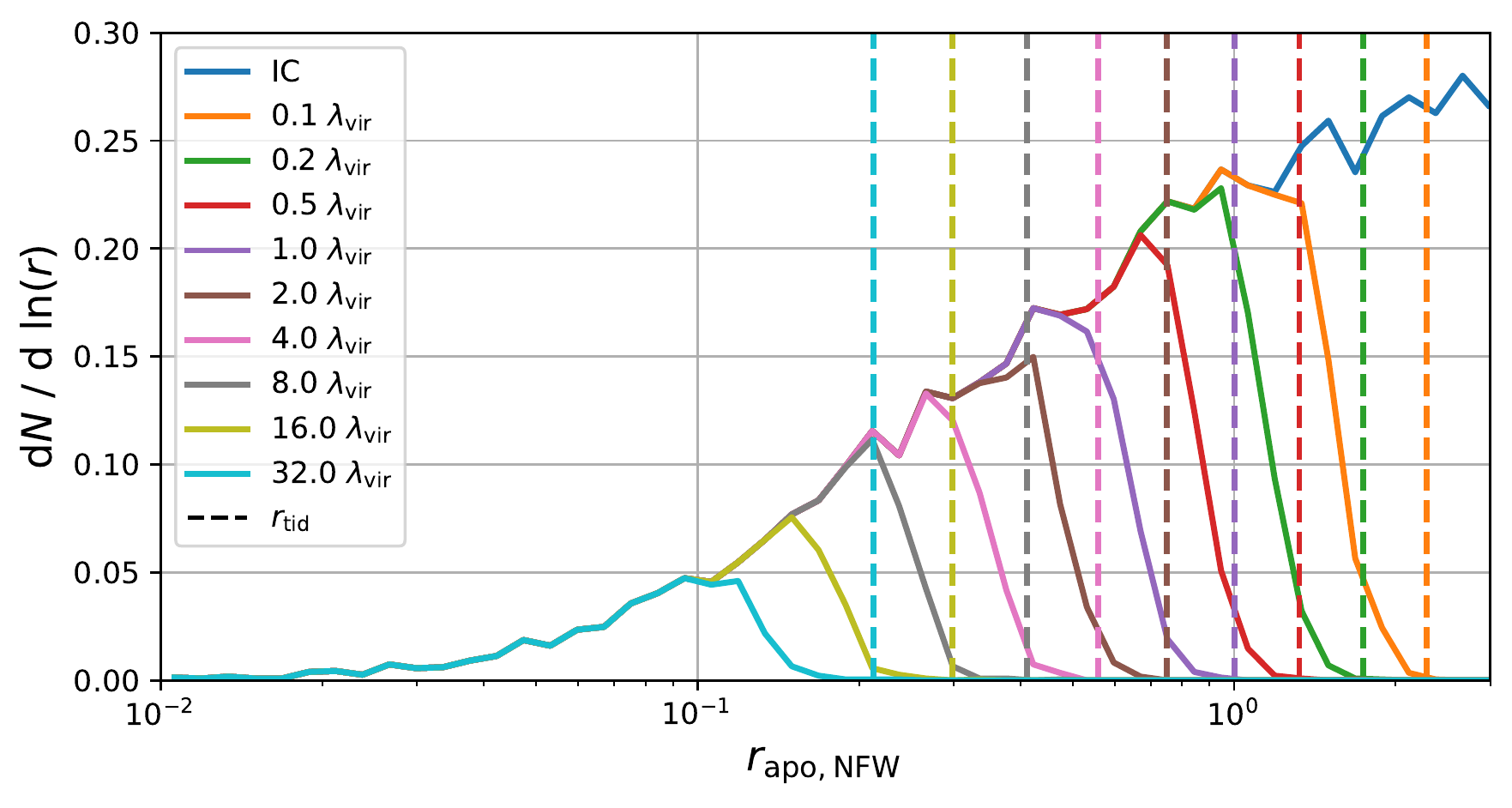}\\
    \includegraphics[width=\columnwidth]{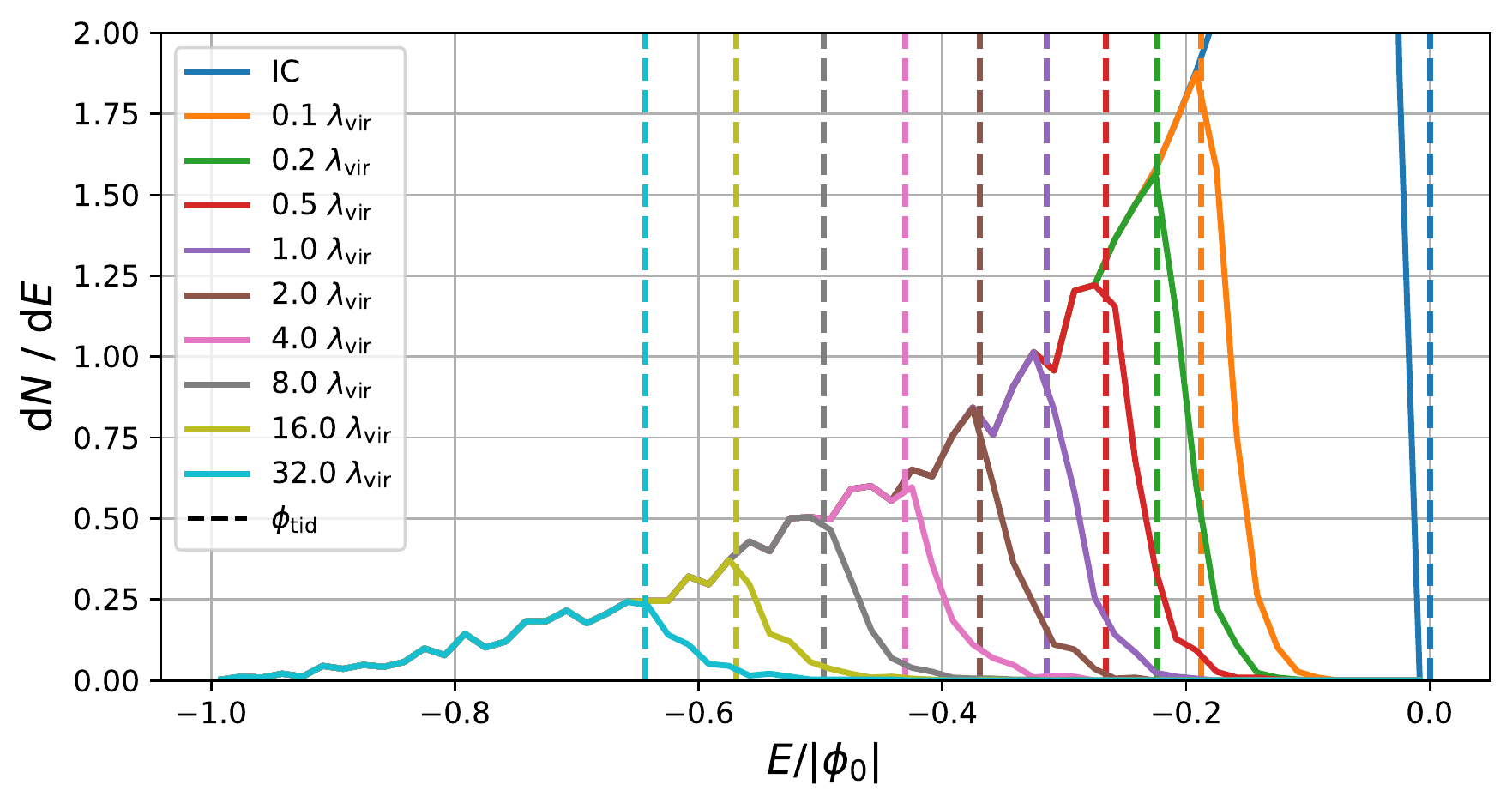}
    \caption{Histograms of particles that remain bound to the system in the adiabatic limit for different amplitudes of the tidal field. Top: apocenter radii with tidal radius marked. Bottom: (Initial) energies with tidal escape energy level marked. It becomes clear that the energy level is more predictive of what particles will leave the system than any criterion based on the tidal radius. The tidal boundary is best understood as an energy level.}
    \label{fig:energysurvival}
\end{figure}

We investigate which properties of the particles in the original NFW profile are most predictive of whether they will escape when the tidal field is applied. \rvtext{Here, we define particles as escaped when they reach a distance of 30 virial radii. (Note that since we are simulating on so long time-scales in the adiabatic limit this value does not matter at all and virtually all particles that exit the potential valley easily reach this radius. All results are robust to choosing smaller or larger escape radii here.)} \rvcom{(Note at the referee: the peculiar value of 30 virial radii here comes from a typo I made when setting up the runs (I intended to use 8 virial radii originally. However, I checked this and the value of this makes no difference whatsoever -- particles which escape the potential valley accelerate exponentially and quickly reach any radius)}

The traditional picture of tidal stripping suggests that particles get stripped when they go beyond the tidal radius. In a first order approximation one could therefore consider all particles  that are initially beyond the tidal radius to escape. However, that would be a very crude approximation, since many particles could easily cross the tidal radius at their next apocenter passage. Therefore, instead it makes more sense to consider the apocenter radii $r_{\rm{apo}}$ (of the self-potential) of the particles. The pericenter and apocenter are given by the two radii which solve the equation
\begin{align}
    E = \frac{L^2}{2 r_{\rm{apo/peri}}^2} + \phi_{\rm{NFW}}(r_{\rm{apo/peri}})
\end{align}
where $E$ is the energy and $L$ the angular momentum of the particle. We infer the apocenters of all particles numerically and plot the distribution of $r_{\rm{apo}}$ for all particles that did not escape the system at the end of the simulation in Figure \ref{fig:energysurvival}. We compare with the tidal radius $r_{\rm{tid}}$ which we define here through the saddle point of the potential
\begin{align}
    \partial_r \phi_{\rm{NFW}} (r_{\rm{tid}}) - \lambda r_{\rm{tid}} &= 0
\end{align}
where we remind the reader that $\lambda$ is the largest eigenvalue of the tidal field. We note that $r_{\rm{apo}} < r_{\rm{tid}}$ is not a very good predictor for particles that stay in the system. I.e. there are a lot of particles that have $r_{\rm{apo}} < r_{\rm{tid}}$ that escape. That is so, since the turn-around radius in the system with applied tidal field can be much larger than the one in the original spherical NFW system, since the tidal field flattens out the potential (in the direction of the largest eigenvalue of the tidal field).

As suggested by the ``deforming bowl'' picture, an alternative predictor of the escaping particles is their energy. The escape energy level is given by the potential in the saddle point
\begin{align}
    \phi_{\rm{tid}} &= \phi_{\rm{self}}(r_{\rm{tid}}) - \frac{1}{2} \lambda r_{\rm{tid}}^2
\end{align}
In the bottom panel of Figure \ref{fig:energysurvival} we show the initial energy of the particles that remain bound to the system. The dashed line marks the saddle point energy level $\phi_{\rm{tid}}$ and different lines show simulations with different amplitudes for the tidal field. In all cases all particles with  $E < \phi_{\rm{tid}}$ remain bound to the system. However, additionally a few particles with larger energies remain bound. \rvtext{We verified that these are indeed not particles which are in the process of escaping. We are not sure why these particles remain in the system} considering that angular momentum is not conserved in the tidal field, but it might be a similar phenomenon to the particles that can remain close to a satellite inside the Jacobi potential of the restricted three body problem as described in \citet{henon_1970}.  That said, the surviving particles with $E > \phi_{\rm{tid}}$ are a rather small population and it is not clear whether they could also stay bound in more complicated time-dependent tidal fields. Even with this inaccuracy, the criterion $E < \phi_{\rm{tid}}$ is an excellent predictor of bound particles.

\subsection{The escape fraction of orbiting subhaloes} \label{sec:orbitsims}

We have seen in the last section that when slowly imposing a tidal field, the criterion $E < \phi_{\rm{tid}}$ is an excellent predictor of the population of particles that will remain bound to a system with fixed self-potential. Here, we want to test whether a similar prediction holds for particles that are bound to a subhalo that orbits around a larger mass halo. In the ``deforming bowl'' picture, we would expect that we can apply the same predictor when using the saddle point energy level that is implied by the tidal field of the host-halo at pericenter $\phi_{\rm{saddle, peri}}$. Since this ``escape energy'' level is only opened briefly during each pericenter passage, we expect that it takes many pericenter passages until all particles with $E > \phi_{\rm{saddle, peri}}$ escape.

We set up simulations where a subhalo with $M_{200c} = 10^6 M_\odot / h$ and concentration $c = 20$ orbits around a milky-way like host halo with $M_{200c,h} = 10^{12} M_\odot / h$ and $c_h = 6$. The potential field is given by
\begin{align}
    \phi(\myvec{x}, t) &= \phi_{\rm{NFW,h}} (\norm{\myvec{x}}) + \phi_{\rm{NFW, s}} (\norm{\myvec{x} - \myvec{x}_{\rm{s}(t)} })
\end{align}
where $\phi_{\rm{NFW,h}}$ and $\phi_{\rm{NFW,s}}$ are the analytic radial NFW potentials corresponding to the host and subhalo and where we integrate the position of the subhalo $\myvec{x}_{\rm{s}}$ numerically in the host potential. We start the subhalo at location $\myvec{x}_s = (r_{\rm{peri}}, 0, 0)^T$ with velocity $\myvec{v}_s = (0, v_{\rm{peri}}, 0)^T$. We consider two sets of simulations. In the first we choose $v_p$ identical to the circular velocity $v_{\rm{h, circ}} (r_{\rm{peri}})$ at that radius so that the subhalo enters a circular orbit at $r_{\rm{peri}}$. In the second set of simulations we choose $v_p = 1.75 v_{\rm{h, circ}} (r_{\rm{peri}})$ so that the subhalo enters an orbit with the same pericenter $r_{\rm{peri}}$, but a much larger apocenter radius.

\begin{figure}
    \centering
    \includegraphics[width=\columnwidth]{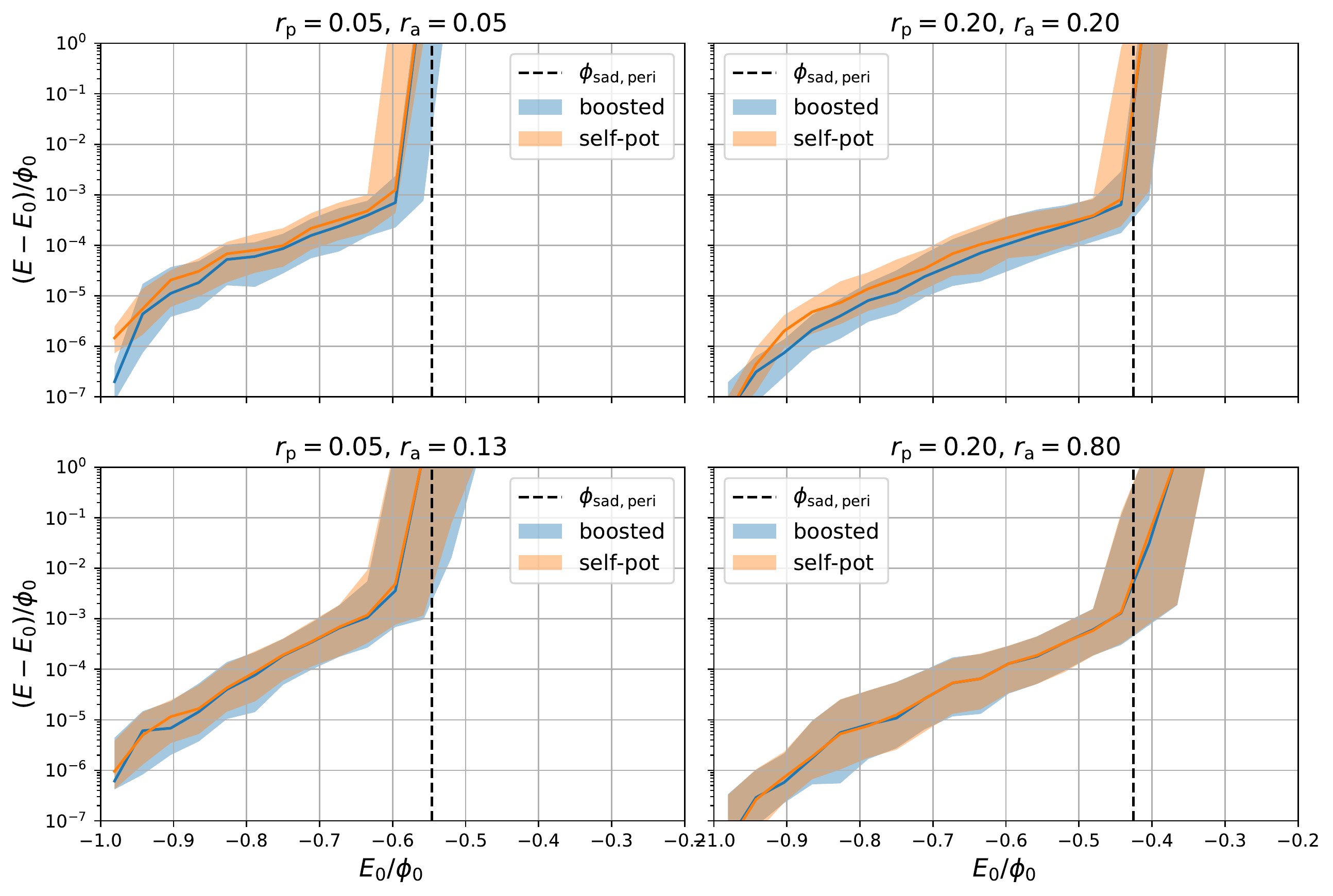}
    \caption{The energy difference of particles between the subhalo's first and 10th apocenter passage. Displayed are the median (solid line) and 20th-80th percentiles (shaded regions). The different panels correspond to subhalos that were started on different orbits -- top: circular orbits and bottom eccentric orbits with the same pericenter. The titles indicate the apo- and pericenters of the orbits in units of the virial radius of the host-halo. It seems that for both, the radial and the circular orbits energies are very well conserved for particles below the pericenter saddle point energy.}
    \label{fig:orbits_energycons}
\end{figure}

First of all, we show that even though the particles orbit in a strongly time-dependent potential, for particles with $E < \phi_{\rm{saddle, peri}}$ the energy level is very close to conserved. In Figure \ref{fig:orbits_energycons} we show the difference between the energies of the particles measured at the 1st and 10th apocenter passage of the subhalo. We define the energy either through the self-potential or through the boosted potential with
\begin{align}
    \phi_{\rm{boost}} &= \phi(\myvec{x}, t) - \nabla \phi_{\rm{NFW,h}}(\myvec{x}_s) \cdot (\myvec{x} - \myvec{x}_s) - \phi_{\rm{NFW,h}} (\myvec{x}_s)
\end{align}
We do not see much difference between the energy definitions on the question of energy conservation here. Importantly, in both cases it seems that energy conservation is dramatically violated for all particles with $E > \phi_{\rm{saddle, peri}}$, because the majority of those particles have escaped the potential valley. On the other hand, the particles with energies $E \lesssim \phi_{\rm{saddle, peri}}$ only change their energy very modestly by less than $10^{-3} \phi_0$ which is quite negligible for our binding criterion. Particles on such orbits are adiabaticaly shielded \citep{weinberg_1994a, weinberg_1994b, Spitzer_1987}. We therefore assume that at first order, \emph{the energy-redistribution through the tidal field can be neglected for particles that remain bound} which is in agreement findings from other studies \citep{vandenbosch_2018}. As already mentioned, energies can still be changed through mass loss which we do not model here, but we will investigate in \citsubhalo{}. 

\begin{figure}
    \centering
    \includegraphics[width=\columnwidth]{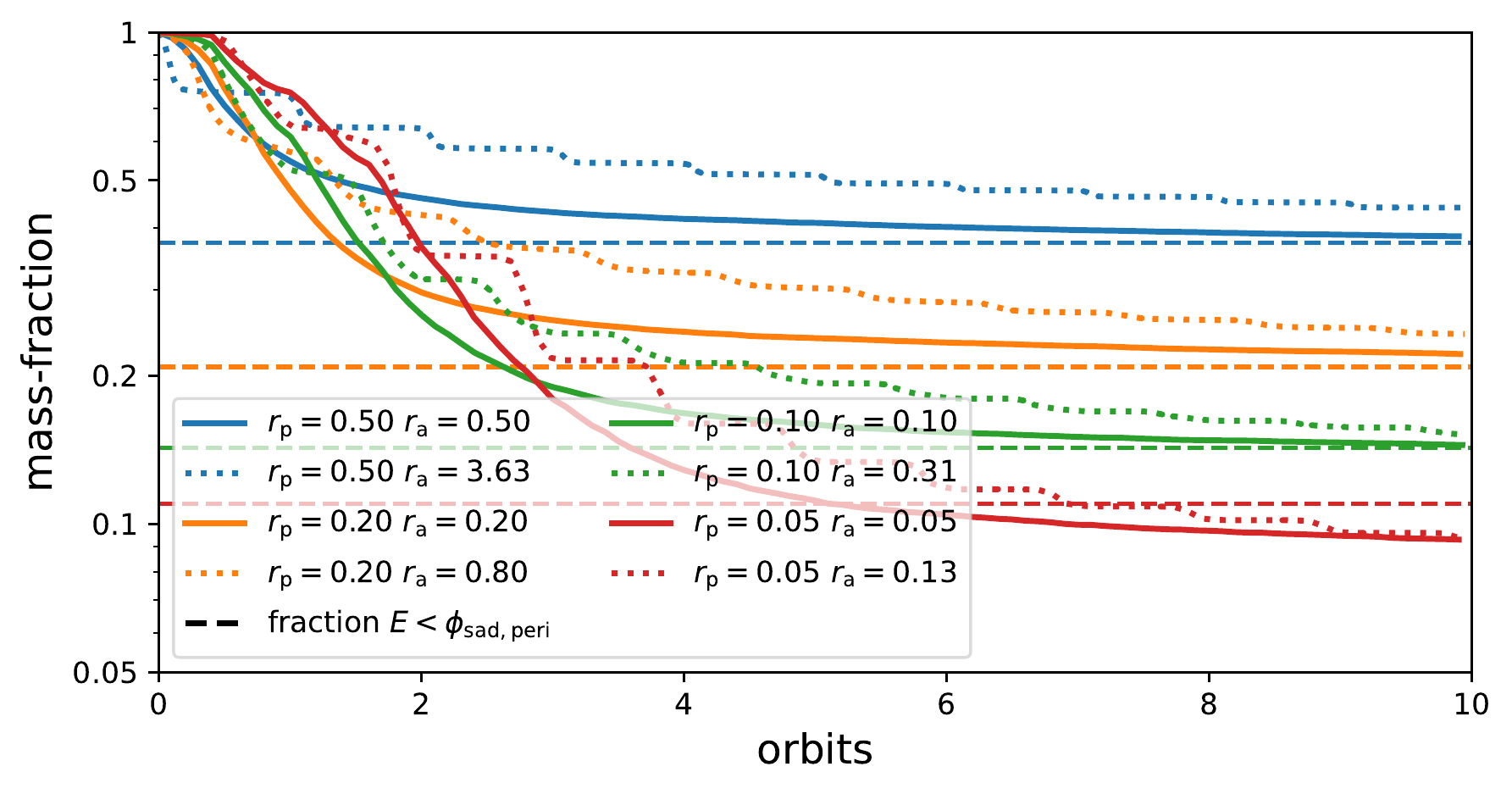}\\
    \includegraphics[width=\columnwidth]{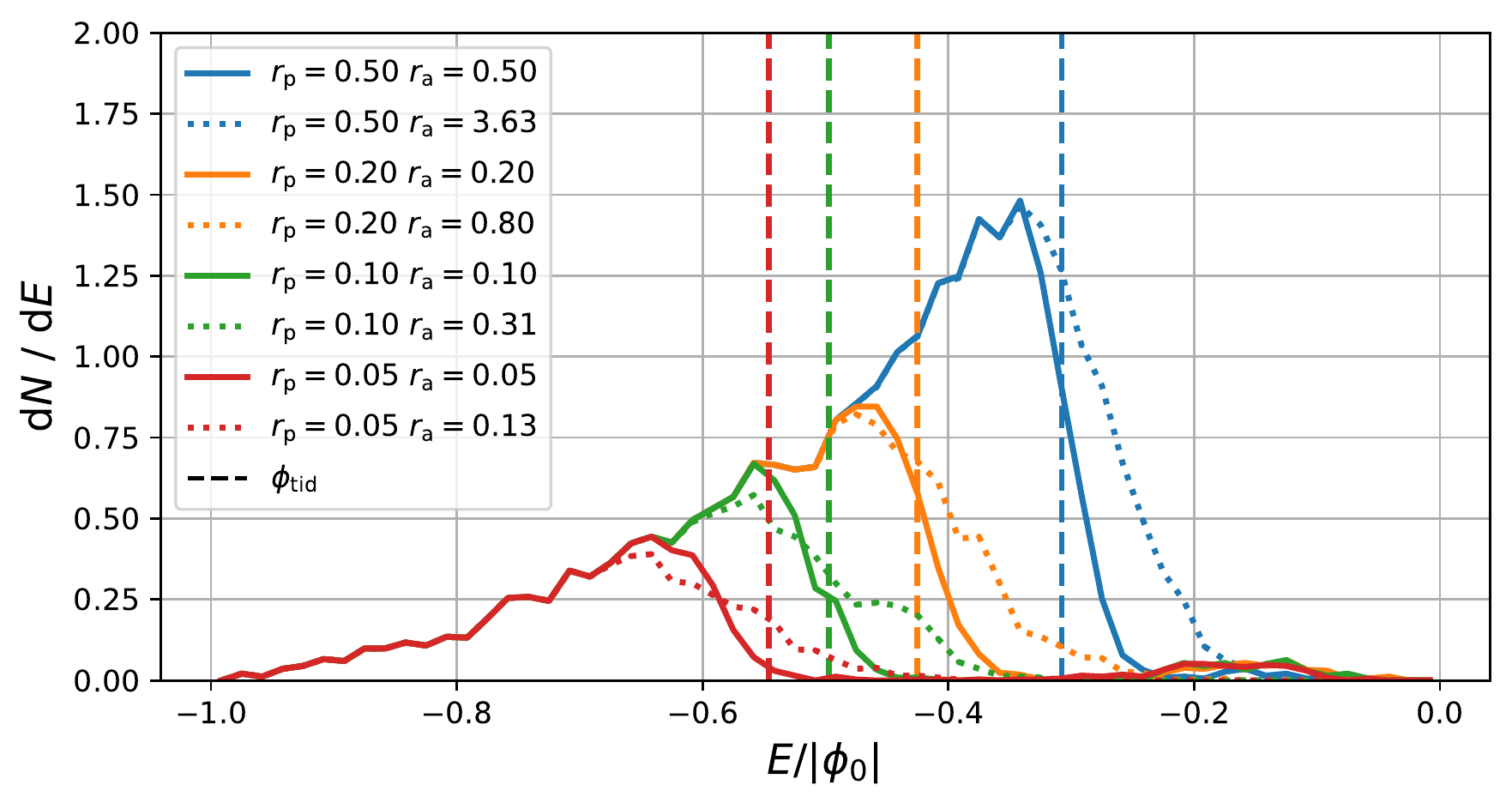}
    \caption{Top: fraction of mass that did not escape the subhalo as a function of the number of (radial) orbits. The solid lines correspond to circular orbits and the dotted lines to eccentric orbits with the same pericenter. The dashed horizontal lines correspond to the bound fractions predicted by $E < \phi_{\rm{saddle, peri}}$. It seems that in the limit of large times (approximately reached after 10 orbits) both the circular and the radial orbiting subhalos reach roughly the same limit that is well approximated by this criterion. Bottom: Distribution of initial self-energies of particles that have remained bound to the subhalo after 10 (radial) orbits. This Figure is the analogue of Figure \ref{fig:energysurvival} and it seems that $E > \phi_{\rm{saddle, peri} }$ is an excellent predictor of which particles will leave the system.}
    \label{fig:orbits_massloss}
\end{figure}

Next we will have a look at the mass loss and whether we can use the criterion $E > \phi_{\rm{saddle, peri}}$ to predict the escaping particles. In the top panel of Figure \ref{fig:orbits_massloss} we show the fraction of mass that did not escape the subhalo as a function of time. We flag a particle as ``escaped'' if it ever goes beyond two times the initial virial radius of the subhalo. Note that we scaled the time by the time needed for one radial period for each case. This orbital period is for the eccentric orbits a factor 2-3 larger than for the circular ones with the same pericenter. That means the circular orbits lose their mass much faster in physical time than radial orbits with the same pericenter. It seems, however, that after a long time (more than 10 orbits) both kinds of orbits converge to an asymptotic limit where no more mass is lost. This limit is well approximated by the criterion $E > \phi_{\rm{saddle, peri}}$. 

In the bottom panel of Figure \ref{fig:orbits_massloss} we show the initial energy distribution of the particles that did not escape. Similar to Figure \ref{fig:energysurvival} we find again that $E > \phi_{\rm{saddle, peri}}$ predicts well which particles are going to leave the system.

\subsection{Discussion}
We have shown that it is possible to get a good approximate prediction of what particles remain bound to a subhalo in the long run just through knowledge of the largest tidal field that the subhalo encounters. The main effect of the tidal field is that it lowers the energy level that is needed to escape the subhalo. This is in qualitative agreement with the measurements of the asymptotic tidal remnants presented in \citet{errani2021} and gives additional motivation for models similar to the energy-space truncation presented in \citet{drakos_2017} and \citet{drakos_2020}.

The ``deforming bowl'' picture of ``tidal overflow'' motivates that one can parametrize the long-term limit of subhalo evolution just through two parameters: (1) the concentration of the subhalo at infall (2) the largest eigenvalue of the tidal field at pericenter. Such a model would be very simple and flexible. For example, it could easily be applied to scenarios with baryonic components as well -- where e.g. a galactic disk would simply enhance the strength of the tidal field. It could be used to calculate lower limits of the mass that resides in low-mass subhaloes and of the self-annihilation signal from very low mass satellite subhaloes.

However, we cannot present such a model here, since it is necessary to account for the revirialization of the subhalo after mass has been lost -- which we explicitly excluded from our considerations here. We think that the most natural way to incorporate it is by consideration of the reaction of the system to an adiabatically applied tidal field. Particles that get lost in this process should be treated as if they were removed infinitely slowly in the adiabatic limit. Such an adiabatic mass loss can be implemented with the recipe that is described in \citet{BinneyTremaine2008}, and we will demonstrate this in a future publication \citsubhalo{}.

Further, we note that most (semi-)analytic approximations of the tidal mass loss of subhaloes model the mass loss rate of subhaloes to be proportional to the mass that resides outside of the instantaneous tidal radius:
\begin{align}
    \dot{m} = - \frac{\alpha}{T_{\rm{orb}}} m(< r_{\rm{tid}})
\end{align}
\citep[for example]{Taylor_2001, vandenbosch_2018, drakos_2020, errani2021}. We speculate that a better criterion might instead consider which fraction of mass has enough energy to escape the system given the instantaneous escape energy level:
\begin{align}
    \dot{m} = -\frac{\beta}{T_{\rm{orb}}} m(E_{\rm{boost}} > \phi_{\rm{tid}})
\end{align}
where $E_{\rm{boost}}$ is the energy associated with the boosted potential, $\phi_{\rm{tid}}$ is the instantaneous escape energy (measured as the saddle point level in the boosted potential) and, possibly, the unknown fore-factor $\beta$ could be determined a priori through theoretical considerations of which fraction of the phase space with $E_{\rm{boost}} > \phi_{\rm{tid}}$ can pass through the saddle point per time.

\section{The existence of boost-invariant topological features} \label{sec:advancedmath}
\begin{figure*}
    \centering
    \includegraphics[width=\textwidth]{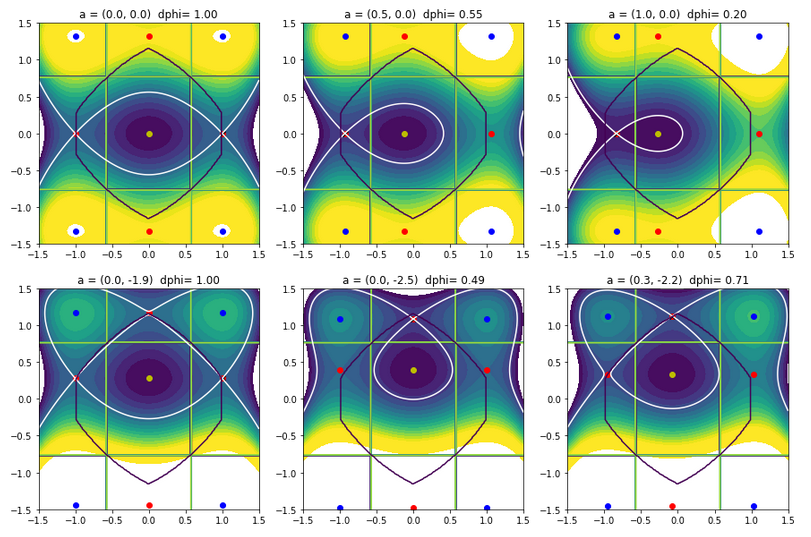}
    \caption{Illustration of a boost-invariant property of a potential field. The different panels show the same potential field with different global gradient terms applied. The critical points (blue: maxima, red: saddle points, yellow: minima) can be found in various different locations (limited by the areas which have the same signs of the Hessian in green) depending on the gradient term. In each system we can construct the region which can ``hold water'' (white contour). The union of all possible water-holding regions is marked as a black contour. This region can be considered an important boost-invariant feature of this ``potential-valley''}
    \label{fig:boostinvariant}
\end{figure*}

So far, we have seen that boost-operations turn the potential into a meaningful quantity in many different contexts -- such as the definition of haloes and binding checks, boundaries of cosmic web structures and the disruption of subhaloes. The main conceptual step that we have taken in this paper, is the realization that a large-scale gradient term in the potential should always be irrelevant for the local dynamics of a system and this has to be considered when interpreting the potential. One simple way to deal with this, is to measure the large-scale gradient term around some location and to construct a boosted potential. However, we have also seen that it is not always straightforward to find a clear and unique choice of such a large-scale gradient, leaving a considerable theoretical degree of freedom of what the right boosted system is.

In this section, we want to briefly discuss the possibility to measure features of the potential field $\phi(\myvec{x})$ that are boost-invariant, i.e. functionals $F$ that are independent of a particular choice of a frame of reference:
\begin{align}
    F(\phi(\myvec{x})) &= F(\phi(\myvec{x}) + \myvec{a} \myvec{x})
\end{align}
While there exist the branches of math ``Morse Theory'' \citep{milnor_1963} and ``Discrete Morse Theory'' \citep{forman_1998, forman_2002, sousbie_2011}  which deal with topological behavior of level sets, critical points and integral lines, we do not know of any more general version of this theory which would consider features that remain invariant under addition of global gradients. Here, we simply show the existence of such features, but hope that mathematicians might in the future develop theories and numerical tools to detect and study them systematically.

One boost-invariant feature of the potential is the maximal region that can hold water under any boost-transformation. To illustrate this, we define the function
\begin{align}
    \phi_a(x, y) &= 2 x^2 + \frac{7}{2} y^2 - x^4 - y^4 + \begin{pmatrix} x \\ y\end{pmatrix} \cdot \myvec{a}
\end{align}
with the vector parameter $\myvec{a}$ that controls the global gradient. We show this function for the case without any boost $\myvec{a} = 0$ in the top left panel of Figure \ref{fig:boostinvariant} and with different boosts in the other panels. In each panel of Figure \ref{fig:boostinvariant} we mark the critical points and the critical level set that corresponds to the persistence-level of the minimum close to $(0,0)^T$. We can imagine that we were holding a strangely shaped bowl where the top of the bowl is anisotropically bent downwards in the x-direction and upwards in the y-direction. The bowl holds some water, and when we turn the bowl (corresponding to boosts), the highest water level changes (indicated as white contours). In each frame the highest water level can be determined by Morse Theory or by watershed algorithms \citep[e.g.][]{Platen_2007}. However, since it depends on the acceleration, each individual water-level is clearly not a boost-invariant feature.

However, we can define the union of all regions that can hold water and this would be a boost-invariant feature of the bowl.  We have indicated this region in black and we have determined it numerically, by boosting into 10000 randomly chosen frames and determining the critical contour in each case. Further, we can define the highest possible water level as a generalization of the persistence. We have indicated the persistence in the title of each of the different frames in Figure \ref{fig:boostinvariant}. We might think of the black region as a boost-invariant generalization of a potential valley. However, whether such a statement makes sense, depends on whether actual physical insights might be gained from this construction. Since, we are lacking a good mathematical understanding of such regions and do not have the numerical tools to construct these ``water-holding'' regions efficiently, we do not attempt any further investigation at this point. However, if an efficient algorithm would be found for constructing these (or similar) boost-invariant features, it might be used to develop a structure finder that operates only on the potential field. 

\section{Outlook and Open Questions}
We have shown in this paper that the potential field of cosmological simulations can be used for a meaningful analysis of the local dynamics if investigated in a manner that is independent of uniform large-scale gradients. The simplest way to create an intuitively understandable potential landscape is to determine such a large-scale gradient close to an object of interest and then to subtract that gradient. Thereby, one switches into a boosted frame of reference in which the potential has locally a much smaller time dependence than in the original frame. We have shown that such a boosted potential can be constructed in many different contexts and we hope to inspire researchers of various fields to incorporate the boosted potential into specific scientific topics.

By using the boosted potential, we have found several striking results in very distinct domains: (1) The boosted potential can be used to include the effect of tidal fields into the binding check of halo-finders. This becomes crucial for reducing misidentifications of unbound overdensties as haloes in warm dark matter simulations. Further, the boosted potential binding check selects truly bound virialized particles of haloes, thereby recovering the expected virial ratio of two on average (unlike other state-of-the art algorithms which are systematically below two). (2) Filaments can also be understood in the language of the boosted potential. The potential landscape around filaments corresponds to a closed potential in the directions orthogonal to the filament and an open unbound potential along the filament axis. In principle, it is possible to define tidal radii and binding checks for filaments. (3) The tidal stripping of subhaloes can be understood from an energy space perspective in the ``deforming bowl'' picture: the mass loss is induced by a lowering of the escape energy level through the tidal field. It predicts that the mass lost in the asymptotic limit depends only on the strength of the highest encountered tidal field (typically at pericenter).

Despite these numerous applications, many details remain to be addressed. One of the main issues is to find efficient methods for defining the gradient to be subtracted. A simple approach to define such a gradient is by calculating local averages over particles. This gives already good results, but it also leaves some considerable degree of freedom in the analysis. Another approach is to consider boost-invariant features of the potential field that might offer unique solutions. Although we are able to show that such boost-invariant features exist, we do not have a rigorous mathematical theory of these. We imagine that they could be handled in some hypothetical generalized version of Morse Theory, and we hope that mathematically more experienced researchers can help to develop such a theory and corresponding numerical methods.

We think that there is sufficient motivation to investigate the boosted potential in future studies. One straight-forward step would be to implement the boosted potential binding check on top of a classical halo finder like e.g. \subfind{} in \textsc{Gadget}. \rvtext{Note that, a good implementation of the boosted potential binding check can be computationally cheaper than \subfind{}, since it only requires a single global potential calculation instead of many individual self-potential computations.} This would make it possible to systematically investigate the effect of tidal fields onto the structure of haloes in large scale simulations. Thereby it could help to understand the boundaries, the triaxiality, the intrinsic alignment and the virialization of haloes. However, in a more general approach that investigates boost-invariant features of the potential field it might also be possible to develop a cosmological structure finder that is solely based on the potential field and that might be able to describe haloes, filaments, and walls in the same language and with meaningful binding criteria.

Further, the boosted potential has motivated us to develop a model of subhalo mass loss in the adiabatic limit which we will present in a forthcoming publication \citsubhalo{}. Finally, we will show in another subsequent paper \citlagrangian{} that the boosted potential can also be constructed in Lagrangian space to understand structure formation from an alternative point of view, and to predict accurately which particles will fall onto which proto-haloes.

\section*{Acknowledgements}

We thank Oliver Hahn, Simon White, Mark Neyrinck, Volker Springel, Go Ogiya and Stephane Colombi for their feedback and helpful comments on the manuscript. J.S. thanks Marcos Pellejero-Ibañez and the BACCO group for interesting and motivating discussions. J.S. and R.A. acknowledge funding from the European Research Council (ERC) under the European Union's Horizon 2020 research and innovation program with grant agreement No. 716151 (BACCO). The authors thankfully acknowledge the computer resources at MareNostrumIV and technical support provided by the Barcelona Supercomputing Center (RES-AECT-2019-3-0015).

\section*{Data Availability}

The data and algorithms underlying this article will be shared on reasonable request to the corresponding author.



\bibliographystyle{mnras}
\bibliography{boosted_pot} 




\appendix

\section{Numerics of the boosted potential binding check} \label{app:boostednumerics}
Here we give a more detailed description of the boosted potential binding check that we have used in Section \ref{sec:bindingcheck}. Our implementation is based on the \TTK library \citep{tierny2018}. The \TTK is an open-source library and software collection for topological data analysis and visualization. We mainly use its discrete Morse theory functions. These are designed to detect critical points and Morse-smale complexes in discrete scalar fields.

(1) For each (sub-)halo, we start with a guess of a (sub-)halo center $\myvec{x}_0$ and the number of particles $N$ that belong to a (sub-)halo. These initial guesses are inferred from the \subfind{} algorithm.

(2) We select a sphere around $\myvec{x}_0$ with a radius $R$ such that it contains $0.8 N$ particles and infer $\myvec{a}_0$ through averaging the accelerations for all particles inside this sphere according to equation \eqref{eqn:boostedaverage}. Note that taking $80\%$ of the particles is somewhat arbitrary, but we have checked that the result of the binding check is reasonably robust to this choice. We use the average in a sphere so that this check does not depend too strongly on the details of the preselection by \subfind{} and to what happens in the outskirts of the object. 

(3) We select a sufficiently large region around this\footnote{sufficiently large means that the region has to be large enough that the saddle point can be properly identified. In practice one can start with a region e.g. containing $2N$ particles and then, if one of the subsequent steps fails (the critical contour touching the boundary of the region), one restarts with a larger selected region.}, evaluate $\phi_{\rm{boost}, i}$ for all particles in this region and perform a Delaunay tessellation of these particles to define a continuous scalar field $\phi_{\rm{boost}} (\myvec{x})$

(4) We us the \TTK code \citep{tierny2018} to find deepest minimum inside of the sphere which contains $0.8N$.

(5) We use \TTK to identify the saddle point where this minimum merges with a deeper minimum. This defines the persistence of the minimum $\Delta \phi = \phi_{\rm{saddle}} - \phi_{\rm{min}}$

(6) We select the set $S_{\rm{i}}$ of all particles which are inside of the critical contour where $\phi_{\rm{boost}} < \phi_{\rm{saddle}}$.

(7) We perform an iterative binding check for this group of particles. We start with $\myvec{v}_0 = \langle \myvec{v} \rangle_{S_i}$ -- the center of mass velocity of all particles that are in $S_i$. We calculate the kinetic energy of each particle $E_{\rm{kin}} = \frac{1}{2} ||\myvec{v} - \myvec{v}_0||^2$ and the total energy $E_{\rm{tot}} = E_{\rm{kin}} + \phi_{\rm{boost}}$ and assign all particles which are in $S_{\rm{i}}$ and which have $E_{\rm{tot}} < \phi_{\rm{saddle}}$ to the set of bound particles $S_{\rm{b}}$. In practice we found this already to work very well when just done once. However, to get the optimal estimate of $\myvec{v}_0$ and therefore $S_{\rm{b}}$ we repeat this step a couple of times where we determine $\myvec{v}_{0}$ for the next iteration from the particles that were in $S_{\rm{b}}$ in the last iteration. 

\section{Energy Conservation and the Boosted Potential} 
\subsection{Tidal fields in the adiabatic limit} \label{sec:adiabaticenergy}
As discussed in Section \ref{sec:sdisruption}, in a time-varying potential the energy is not conserved. Therefore, also in our experiments where we slowly apply a tidal field to an NFW halo, energy is formally not conserved. However, we will show here briefly, (1) that energy conservation is violated so little that it is hardly relevant for the proposed binding checks. (2) Energy conservation is less strongly violated in the adiabatic limit when considering the boosted potential instead of the self-potential.

\begin{figure}
    \centering
    \includegraphics[width=\columnwidth]{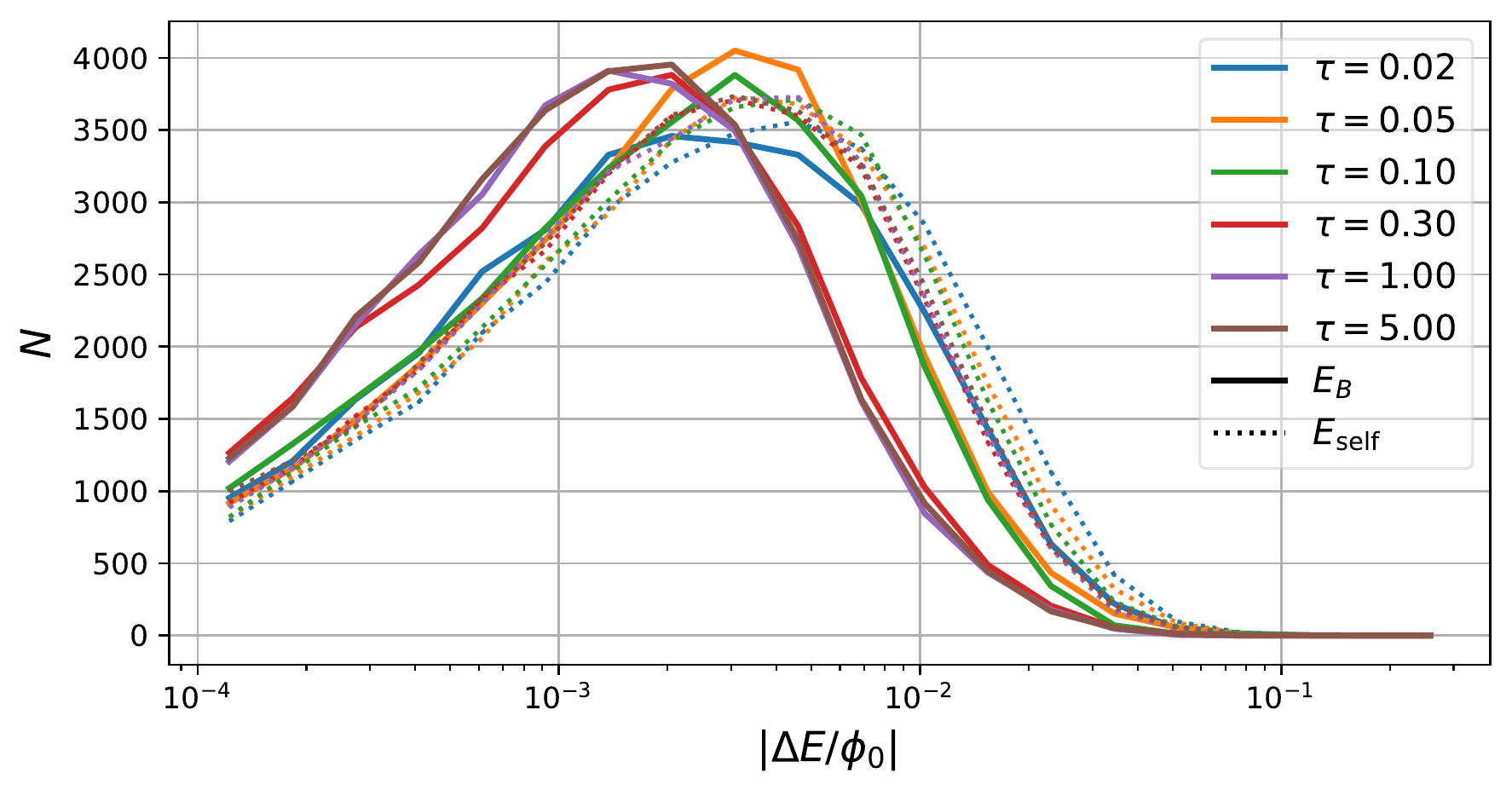}
    \caption{Energy Conservation for an NFW haloes with a slowly applied tidal field. The histogram is taken over all particles that stay in the system in the long run and different colours correspond to tidal fields that were increased at different rates, but with the same final amplitude. $\tau \rightarrow \infty$ corresponds to the adiabatic limit. Solid lines correspond to the energy defined by the boosted-potential and dotted lines to the energy defined through the self-potential. The boosted energy conservation is less violated in the adiabatic limit. However, for both potential definitions typical changes in energy are quite negligible in comparison to the binding energy $\Delta E \ll 0.1 |\phi_0|$.}
    \label{fig:energyadiabticlimit}
\end{figure}

In Figure \ref{fig:energyadiabticlimit} we show the energy distribution of particles that remained bound to the system with a final tidal field of $8 \lambda_{\rm{vir}}$ in the long-time limit. Different colours correspond to simulations where the tidal field was applied at different rates and $\tau \rightarrow \infty$ would correspond to the adiabatic limit. We can see that that the boosted energy conservation is less violated in the adiabatic limit than for more instantaneous changes. We imagine that this is so, since a (trace-free) tidal field lowers the potential energy of a spherical orbit on average as much as it increases it. In the adiabatic limit, the time-dependent potential effectively interacts with the whole orbit of each particle simultaneously, thereby roughly balancing out the positive and negative energy change. On the other hand, the self-potential neglects the tidal component in the potential and therefore particles' energies keep oscillating always in the presence of the tidal field.

However, we note that in both cases, the change in energies of these particles is quite negligible when compared to the maximal binding energy $|\phi_0|$. The largest changes correspond to $10 \%$ of this binding energy, but more typical changes are less than $1\%$. Neither of these changes are very dramatic if thought of as a perturbation on the x-axis of Figure \ref{fig:energysurvival}.

\subsection{Energy redistribution through tidal fields}
\begin{figure*}
    \centering
    \includegraphics[width=\textwidth]{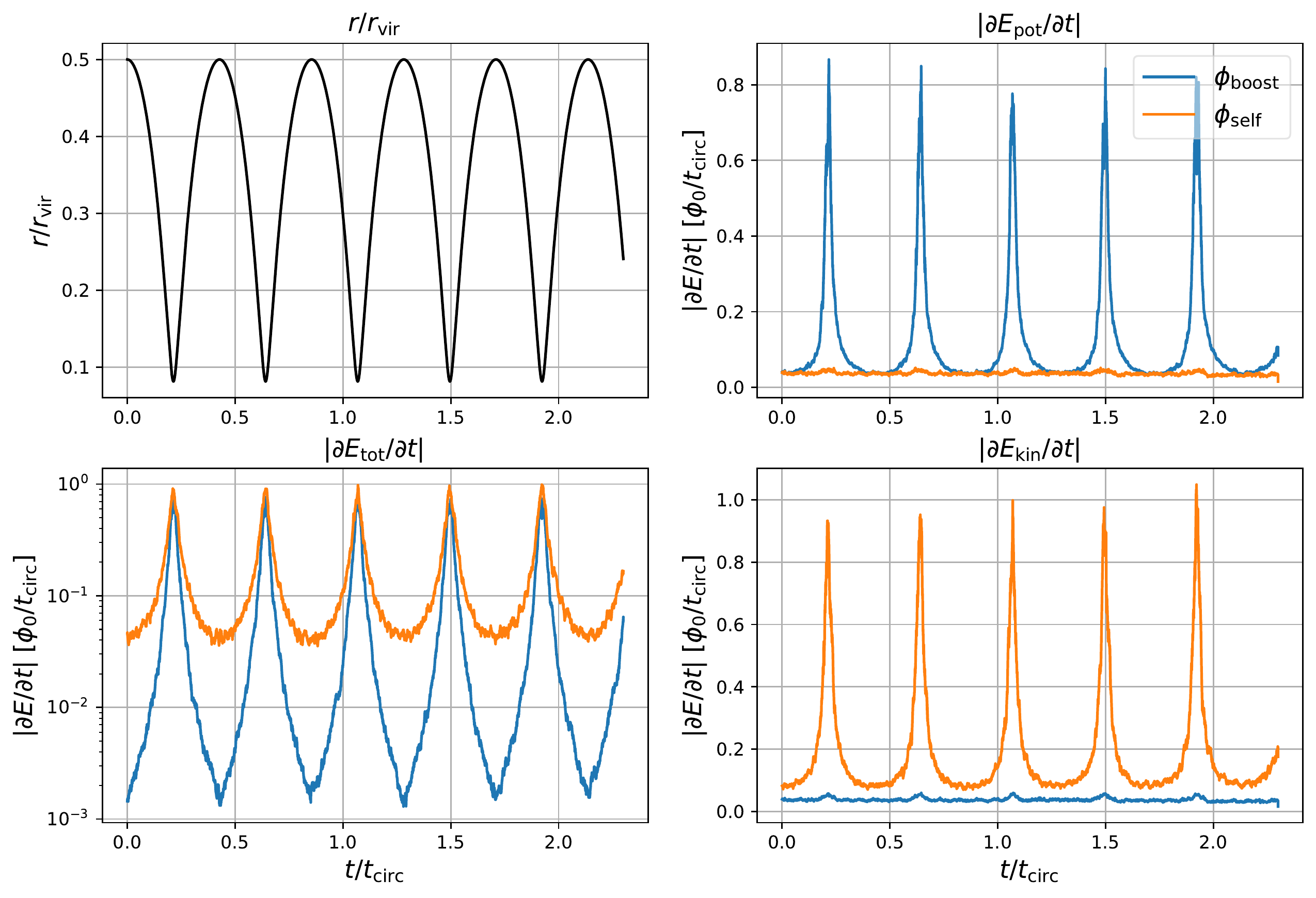}
    \caption{Energy-injection for an orbiting subhalo. The top-left panel shows the radial location of the subhalo as a function of time, the other panels show the median instantaneous rates of energy injection as defined in equations \eqref{eqn:einjkin} - \eqref{eqn:einjtot}. The total rate of energy injection (bottom left) is highest at the pericenter for both energy-definitions. However, for the self-potential the total energy injection rate at apocenter appears more than an order of magnitude higher than for the boosted-potential. This is so, since the self-potential energy of particles keeps oscillating, even if the tidal field changes very slowly (or not at all). For the boosted potential the energy injection appears to happen almost instantaneously at the pericenter through the potential (top right panel), whereas for the self-potential energy injection appears to happen kinetically (bottom right panel).}
    \label{fig:energyinjection}
\end{figure*}

We briefly want to show here that the boosted potential and the self-potential offer quite different perspectives on the question of \emph{how} energy is injected into an orbiting subhalo. From the perspective of the self-potential, energy appears to be injected kinetically through a constantly changing tidal field. Energy seems to be redistributed at all times, but the rate at which it is redistributed is highest in the pericenter where the tidal field is the strongest. This is typically summarized under the notion ``kinetic heating''.

While there is nothing wrong with this perspective, we want to show here that energy injection can be understood quite differently in terms of the boosted potential. As we have seen in \ref{sec:adiabaticenergy}, the boosted energy changes almost not at all in slowly changing tidal fields, but only in the case of very rapid changes in the tidal field. When a subhalo passes through pericenter the tidal field flips almost instantenously its orientation. Therefore, the boosted potential changes instantaneously as well -- particles are lifted or lowered to a new energy level while kinetic energies have no time to react. Subsequently particles orbit in the newly set potential landscape with their new energy. The energy appears to be redistributed ``potentially'' in this perspective.

We  demonstrate this quantitatively in Figure \ref{fig:energyinjection} for a subhalo with concentration $10$, on a very eccentric orbit with apocenter at $0.5 r_{\rm{vir}}$ and pericenter at $0.07 r_{\rm{vir}}$. We define the instantaneous rates of energy injection for a particle as
\begin{align}
    \frac{\partial E_{\rm{kin}}}{\partial t} &= \frac{\rm{d} E_{\rm{kin}} }{\rm{d} t} + \myvec{v} \cdot \myvec{\nabla} \phi \label{eqn:einjkin} \\
    \frac{\partial E_{\rm{pot}}}{\partial t} &= \frac{\rm{d} E_{\rm{pot}} }{\rm{d} t} - \myvec{v} \cdot \myvec{\nabla} \phi \\
    \frac{\partial E_{\rm{tot}}}{\partial t} &= \frac{\partial E_{\rm{kin}}}{\partial t} + \frac{\partial E_{\rm{pot}}}{\partial t} \label{eqn:einjtot} 
\end{align}
where the gradient of the potential enters in the kinetic energy change as the rate at which velocities change, whereas it enters in the potential through the expected energy change through movement to a new potential location.
In a conservative potential both of these terms would be zero. In our simulation we measure $d E^* / dt$ by finite-differences on the corresponding energies of particles between subsequent time-steps and all the other quantities can be measured instantaneously. For the potential $\phi$ we use the boosted and the self-potential respectively. We show the thus measured instantaneous rates of energy injection in Figure \ref{fig:energyinjection}. The lines represent the median over all particles that remained inside the subhalo until the end of the simulation.

It becomes clear that in the boosted-potential the energy injection happens through the instantaneous lifting or lowering of the potential landscape $\partial \phi / \partial t$. This might make it easier to track such energy changes explicitly. Further, we notice that the boosted energy appears to be much better conserved in the apocenter where the tidal field is changing slowly. This is so, since the self-potential energy of particles is oscillating when they orbit in a constant tidal field whereas the boosted potential energy is exactly conserved if a tidal field is held constant. We speculate that energy injection might be easily understood and tracked through the integrated change in potential energy in the boosted frame
\begin{align}
    \Delta E &= \int \frac{\partial \phiboost(\myvec{x}(t), t)}{\partial t} \rm{d}t
\end{align}
However, as we have argued in Section \ref{sec:orbitsims}, we think that the actual level of energy redistribution through the tidal field is so low for particles that remain bound to an orbiting subhalo, that one might neglect it at first order all-together.


\bsp	
\label{lastpage}
\end{document}